\documentclass[letterpaper, paper,10pt]{AAS}		
\usepackage{overcite}
\usepackage{footnpag}			      	

\usepackage{placeins}
\usepackage{enumitem}

\usepackage[super]{nth}

\usepackage{amsmath,amssymb,amsfonts,amsthm,bm}
    \newcommand{\mb}{\mathbf}
    
    \makeatletter
    \renewcommand{\dddot}[1]{%
        {\mathop{#1\hspace{0pt}}\limits^{\vbox to-1.4\ex@{\kern-\tw@\ex@
                    \hbox {\normalfont .\kern-.1em.\kern-.1em.}\vss}}}}
    \renewcommand{\ddddot}[1]{%
        {\mathop{#1\hspace{0pt}}\limits^{\vbox to-1.4\ex@{\kern-\tw@\ex@
                    \hbox{\normalfont\clap{.\kern-.1em.\kern-.1em.\kern-.1em.}}\vss}}}}
    \makeatother
\usepackage{mathtools}

    \DeclarePairedDelimiter{\floor}\lfloor\rfloor
    \DeclarePairedDelimiter{\abs} \lvert\rvert
    \DeclarePairedDelimiter{\norm}\lVert\rVert
\usepackage{optidef}
\usepackage{tensor}
\usepackage{array}
    \newcolumntype{L}{>{$}l<{$}} 
    \newcolumntype{R}{>{$}r<{$}} 
    \newcolumntype{C}{>{$}c<{$}} 

\DeclareMathOperator*{\+}{\hphantom{-}}
\let\-\relax
\DeclareMathOperator*{\-}{{-}}

\usepackage{siunitx}

\usepackage{graphicx}
\usepackage[labelfont=bf,textfont=bf]{caption}
\usepackage{subcaption}
\usepackage{wrapfig}
\usepackage{rotating}
\usepackage{multirow}

\usepackage{nameref}
    
\usepackage{hyperref}
    \hypersetup{
      linkcolor    = black,
      citecolor    = black,
      urlcolor     = black,
      colorlinks   = true,
      pdfnewwindow = true,
      pdfstartview = FitV,
    }
\usepackage{doi}
\usepackage{listings}

\graphicspath{{./figures/}}

\PaperNumber{22-560}

\begin{document}

\title{A Norm-Minimization Algorithm for Solving the Cold-Start Problem with XNAV}

\author{
    Linyi Hou\thanks{Graduate Research Assistant, Department of Aerospace Engineering, University of Illinois at Urbana-Champaign}{ }
    and 
    Zachary R. Putnam\thanks{Assistant Professor, Department of Aerospace Engineering, University of Illinois at Urbana-Champaign}
}

\maketitle{}

\begin{abstract}
An algorithm is presented for solving the cold-start problem using observations of X-ray pulsars. Using a norm-minimization-based approach, the algorithm extends Lohan's banded-error intersection model to 3-dimensional space while reducing compute time by an order of magnitude. Higher-fidelity X-ray pulsar signal models, including the parallax effect, Shapiro delay, time dilation, and higher-order pulsar timing models, are considered. The feasibility of solving the cold-start problem using X-ray pulsar navigation is revisited with the improved models and prior knowledge requirements are discussed. Monte Carlo simulations are used to establish upper bounds on uncertainty and determine the accuracy of the algorithm. 
Results indicate that it is necessary to account for the parallax effect, time dilation, and higher-order pulsar timing models in order to successfully determine the position of the spacecraft in a cold-start scenario. The algorithm can uniquely identify a candidate spacecraft position within a 10 AU $\times$ 10 AU $\times$ 0.01 AU spheroid domain by observing eight to nine pulsars. The median position error of the algorithm is on the order of 15 km. Prior knowledge of spacecraft position is technically required, but only to an accuracy of 100 AU, making it practically unnecessary for navigation within the Solar System. Results further indicate that choosing lower frequency pulsars increases the maximum domain size but also increases position error. 
\end{abstract}
\section{Introduction}
\label{sec:introduction}

Pulsars emit periodic electromagnetic signals known for their remarkable frequency stability that may rival that of an atomic clock \cite{allan1987millisecond}. X-ray pulsar navigation (XNAV) leverages the timing stability of pulsars to perform navigation in space by measuring the phases of one or more pulsar signals. Most existing XNAV techniques provide updates relative to known a spacecraft state, and have been shown to provide improved state estimate accuracy relative to the Deep Space Network for navigation beyond 15 AU from Earth \cite{graven2008xnav}. Recently, experiments onboard the International Space Station have also demonstrated XNAV capabilities for relative position updates in low Earth orbit \cite{winternitz2016sextant}. 

The cold-start, or lost-in-space, problem is the task of determining the position of a spacecraft (in some known reference frame) with no prior state information and with no external communication. Existing solutions to the cold-start problem are scarce and come with compromises, such as assuming the existence of artificial beacons \cite{tanygin2014closed}, requiring proximity to Earth and the moon \cite{adams2017lost}, assuming knowledge of prior state and the ability to estimate distance to the Sun \cite{dahir2020lost}, or being highly sensitive to the trajectory and measurement conditions \cite{hollenberg2020geometric,hou2021compare}. Sheikh was the first to envision the use of XNAV for solving the cold-start problem with only an onboard clock and stored pulsar models\cite{sheikhUseVariableCelestial2005}. More recently, Lohan developed the first XNAV algorithm to solve the cold-start problem in two dimensions, with some limited analyses in three dimensions\cite{lohan2021methodology}. 

\begin{figure}[ht!]
    \centering
    \begin{minipage}{0.45\linewidth}
        \centering
        \includegraphics[width=0.7\textwidth]{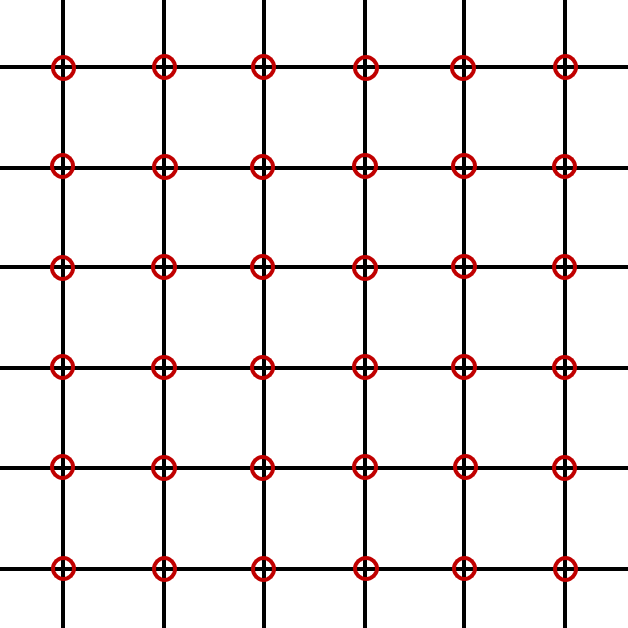}
        \subcaption{}
    \end{minipage}
    \begin{minipage}{0.45\linewidth}
        \centering
        \includegraphics[width=0.7\textwidth]{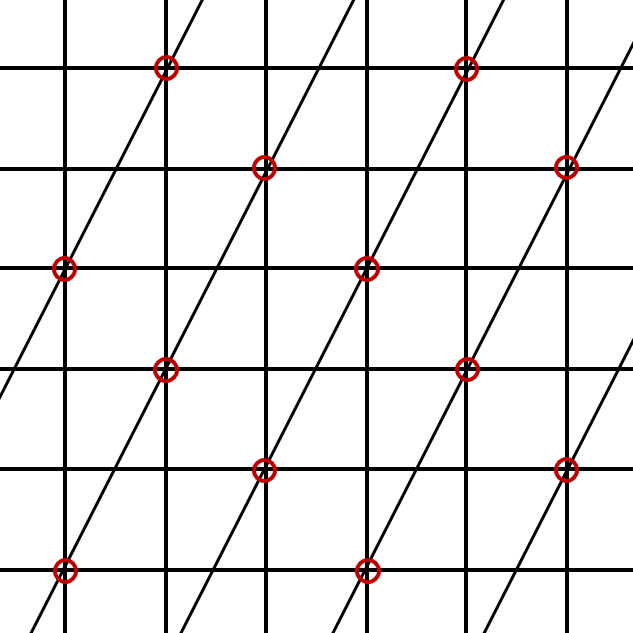}
        \subcaption{}
    \end{minipage}
    \caption{Candidate points from intersecting wavefronts of two pulsars (a) and three puslars (b)\cite{lohan2021methodology}.}
    \label{fig:intersection_model}
\end{figure}

The key concept for using XNAV to solve the cold-start problem is as follows: the phase of a pulsar's signal may be determined from observations, and the observer can be on any wavefront whose phase is equal to the observed value. By observing additional pulsars, the observer's position can be further restricted to the points that lie at the intersection of wavefronts from each observed pulsar. The true position of the observer can be found by observing a sufficient number of pulsars such that only a single intersection point remains within a given spatial domain of finite size. Figure~\ref{fig:intersection_model} illustrates this concept in two dimensions.

\begin{figure}[ht!]
    \centering
    \includegraphics[width=0.7\linewidth]{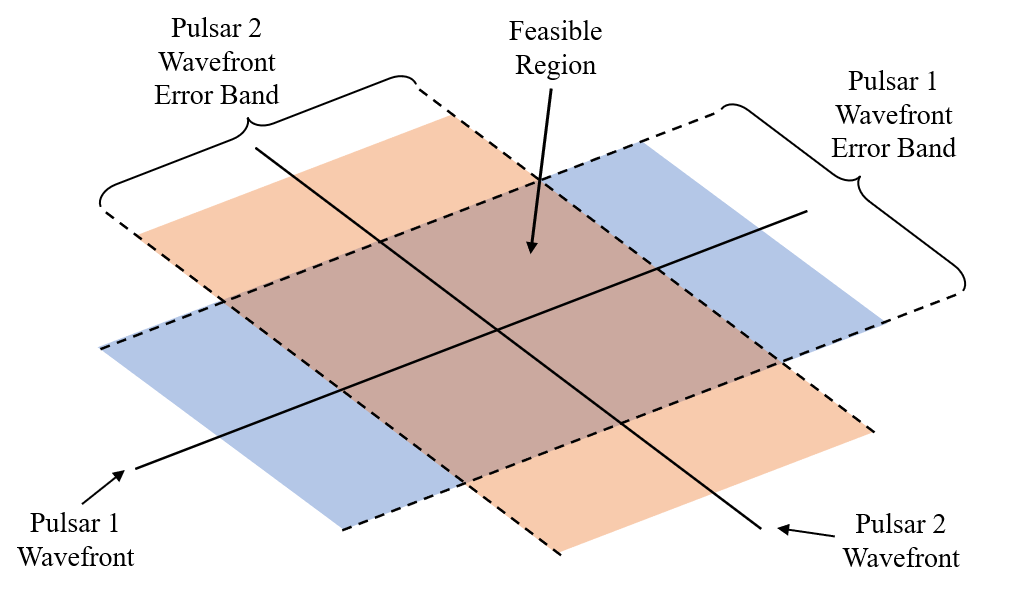}
    \caption{Banded-error intersection model\cite{lohan2021methodology}.}
    \label{fig:banded_error_model}
\end{figure}

The algorithm developed by Lohan established the concept of a banded-error model. Since there is some uncertainty in the phase of each observed pulsar, there is also uncertainty in the position of the wavefronts, which forms a finite region around each wavefront. Rather than singular points, the intersections of wavefronts from multiple pulsars instead identify candidate regions where the spacecraft may be, as illustrated in Figure~\ref{fig:banded_error_model}. In Lohan's work, these regions were found by identifying and tracking the vertices generated by the intersecting error bands. With a suitable choice of pulsars and a bounded spatial domain, case studies showed that a unique candidate region could be found that contained the true position\cite{lohan2021methodology}.

While Lohan's algorithm and analysis could be extended to three dimensions in theory, the algorithm is computationally expensive to run due to the number of vertices formed by intersecting banded regions in three dimensions. Furthermore, Lohan's analysis does not account for key physical aspects of pulsar signals, such as the parallax effect and pulsar frequency variations. The XNAV cold-start algorithm discussed in this paper addresses these shortcomings and revisits the performance capabilities of XNAV in a cold-start scenario with greater scrutiny. 

\section{Signal and Observation Models}
\label{sec:models}

In this section, the equations used to model pulsar observations are described. In contrast to Reference~\citenum{lohan2021methodology}, which only modeled the first-order Doppler delay, the present algorithm also accounts for wavefront time-of-arrival (TOA) delays caused by the parallax effect, the Shapiro delay, time dilation, and the variation in pulsar frequency. These time delays are used to compute the position of a pulsar wavefront based on the speed of light. A later section explores whether these effects are critical to the algorithm's ability to determine spacecraft position.

The notation used in this section is consistent with that of Reference~\citenum{sheikh2007high}. Let $\mb{b}$ be the position of a reference point in an assumed inertial frame centered at the solar system barycenter (SSB), and let $\mb{p}$ be the position of an observer (typically the spacecraft of interest) in the same inertial frame. The relative position of the observer with respect to the reference point, $\mb{r}$, is then $\mb{r} = \mb{p}-\mb{b}$. Furthermore, let $\hat{\mb{n}}$ be a unit vector pointing from the reference point to a pulsar. Assume that the pulsar is sufficiently far from the origin, reference point, and observer, such that the unit vectors pointing from these positions to the pulsar are all essentially the same, i.e., $\hat{\mb{n}}_b \approx \hat{\mb{n}}_p \approx \hat{\mb{n}}_r \approx \hat{\mb{n}}$. According to the Australia Telescope National Facility (ATNF) Pulsar Catalog, the pulsar with the closest estimated distance is B1055-52 at over 300 light-years ($1.92\times10^7$ AU) from the Solar System \cite{ATNFPulsarCatalogue}. Even if $\mb{b}$ and $\mb{p}$ are 100 AU away, their respective pulsar vectors $\hat{\mb{n}}_b$ and $\hat{\mb{n}}_p$ would only have an angular difference of \SI{5.21e-6} rad.

\subsection{Roemer Delay}

The Roemer delay is composed of the first-order Doppler delay and the parallax effect, which describe the time of flight of photons between two points \cite{huygens1899oeuvres,edwards2006tempo2,sheikh2007high,backer1986pulsar}. 

\subsubsection{First-Order Doppler Delay.}
The first-order Doppler delay describes the photon time of flight between two points in space due to the finite speed of light, assuming no curvature of spacetime. Furthermore, the pulsar wavefront is assumed to be planar, i.e., the pulsar is at an infinite distance such that the curvature of each spherical pulse is ignored. Let $t_{\mb{p}}$ and $t_{\mb{b}}$ be the times of arrival of the same wavefront at $\mb{p}$ and $\mb{b}$, respectively. Let $c$ be the speed of light in a vacuum. The first-order Doppler delay is as follows:
\begin{equation}
    \label{eqn:doppler_delay}
    \big(t_{\mb{b}} - t_\mb{p}\big) = \dfrac{\hat{\mb{n}}\cdot{}\mb{r}}{c}
\end{equation}

\subsubsection{Parallax Effect.}
Rather than assuming that pulsars are infinitely far away and that their wavefronts can be approximated as planes, the parallax effect accounts for the spherical geometry of the wavefronts emitted by a pulsar. Let a pulsar be at distance $D$ from the reference point $\mb{b}$; assume the pulsar is sufficiently far from the Solar System such that its distance to different points within the Solar System can be considered the same, e.g., $D_\mb{b}\approx{}D_\mb{p}\approx{}D$. Recall also that $\hat{\mb{n}}_b \approx \hat{\mb{n}}_p \approx \hat{\mb{n}}$. A simplified equation for the parallax effect is considered, which neglects terms of order $\mathcal{O}(1/D^2)$ and higher. Proper motion, which is the apparent movement of stars in the sky, is assumed to be negligible compared to $D$. The parallax effect accounts for the difference in the TOA of the same pulsar wavefront observed at $\mb{b}$ and $\mb{p}$:
\begin{equation}
    \label{eqn:parallax_effect}
    \big(t_{\mb{b}} - t_\mb{p}\big) = 
    \dfrac{1}{2cD}
    \Big[ (\hat{\mb{n}}\cdot{}\mb{r})^2 
        - r^2 
        + 2(\hat{\mb{n}}\cdot{}\mb{b})(\hat{\mb{n}}\cdot{}\mb{r})
        - 2(\mb{b}\cdot{}\mb{r})\Big]
\end{equation}

The Roemer delay is the combination of the parallax effect and the first-order Doppler delay:
\begin{equation}
    \label{eqn:roemer_delay}
    \big(t_{\mb{b}} - t_\mb{p}\big)_{_{R\bm{_\odot}}} = 
    \dfrac{\hat{\mb{n}}\cdot{}\mb{r}}{c} + 
    \dfrac{1}{2cD}
    \Big[ (\hat{\mb{n}}\cdot{}\mb{r})^2 
    - r^2 
    + 2(\hat{\mb{n}}\cdot{}\mb{b})(\hat{\mb{n}}\cdot{}\mb{r})
    - 2(\mb{b}\cdot{}\mb{r})\Big]
\end{equation}

\subsection{Shapiro Delay}

The Shapiro delay describes the bending of photon trajectories near a gravitational field; all masses near and along the photon's trajectory may influence the photon time of arrival \cite{edwards2006tempo2,sheikh2007high,backer1986pulsar}. The present analysis considers only the influence of the Sun within the Solar System while assuming the contributions from other sources to be constant or negligible. Let the observer position and reference point position, $\mb{p}_{_k}$ and $\mb{b}_{_k}$, be defined relative to the k\textsuperscript{th} body in the solar system with $\mb{r}_k = \mb{p}_k - \mb{b}_k$. Denoting the set of relevant celestial bodies in the Solar System as $B_{SS}$, the equation for the Shapiro delay between the two points is as follows:
\begin{equation}
    \label{eqn:shapiro_delay}
    \big(t_{\mb{b}} - t_\mb{p}\big)_{_{S\bm{_\odot}}} = \sum_{k}^{B_{SS}}
    \dfrac{2GM_k}{c^3}
    \ln{\bigg\lvert
        \dfrac{\hat{\mb{n}}\cdot\mb{p}_{_k}+p_{_k}}
              {\hat{\mb{n}}\cdot\mb{b}_{_k}+b_{_k}}
        \bigg\rvert}
\end{equation}

\subsection{Time Dilation}

Time dilation occurs when two clocks are moving at different velocities or operating in different gravitational fields. For XNAV, the spacecraft's onboard clock measures proper time $\tau$ and experiences time dilation with respect to a reference clock in the SSB-centered inertial frame, which measures coordinate time $t$. Terrestrial Time or Barycentric Coordinate Time are typically used in conjunction with the SSB-centered inertial frame \cite{desvignes2016high,freire2012relativistic,ATNFPulsarCatalogue}.

We assume that proper time $\tau$ was synchronized with coordinate time $t$ at some time $\tau_0$, such that $\tau_0 = t_0$. At a later time $\tau_f$, the time dilation can be computed by integration as shown in Eq.~\eqref{eqn:time_dilation}, where $U$ is the total gravitational potential acting on the spacecraft, $r_k$ is the distance between the spacecraft, and the $k$\textsuperscript{th} body as defined in Eq.~\eqref{eqn:grav_potential} \cite{thomas1975reformulation,sheikhUseVariableCelestial2005}.
\begin{equation}
\label{eqn:time_dilation}
    (t_f-t_0) = (\tau_f-\tau_0) + \int_{\tau_0}^{\tau_f}
    \bigg[\dfrac{U}{c^2}+\dfrac{1}{2}\Big(\dfrac{v}{c}\Big)^2\bigg]\;d\tau
\end{equation}
\begin{equation}
\label{eqn:grav_potential}
    U = \sum_{k}^{B_{SS}} \dfrac{GM_k}{r_k} + \mathcal{O}(r_k^2)
\end{equation}

If only the Sun's influence is included, assuming that the spacecraft is in a heliocentric orbit, the integral can be directly evaluated given the semi-major axis $a$ and the starting and ending eccentric anomalies $E_0$ and $E_f$, respectively\cite{moyer1981transformation,sheikhUseVariableCelestial2005}:
\begin{equation}
    (t_f-t_0) = (\tau_f-\tau_0)(1-\mu_s/2c^2a) + (2/c^2)\sqrt{a\mu_s}(E_f-E_0)
\end{equation}

\subsection{Pulsar Timing Model}

Pulsar timing models are defined by a reference time and a reference observatory position, commonly based on the coordinate time in the SSB-centered inertial frame with the SSB as the reference observatory \cite{ATNFPulsarCatalogue}. Given a time $t_0$ and position $\mb{b}$, the phase of a pulsar at time $t_f$ can be characterized by a series of frequency derivatives, $f_{t_0},\dot{f}_{t_0},\ddot{f}_{t_0},\cdots$, which form a Taylor series expansion around $t_0$ as shown in Eq.~\eqref{eqn:phase_original}. $\phi_{\mb{b},t_0}$ denotes the pulsar phase measured at position $\mb{b}$ at time $t_0$, and $\Phi_{\mb{b},t_0}(t_f)$ denotes the Taylor series used to compute the pulsar phase at time $t_f$ at position $\mb{b}$ as shown in Eq.~\eqref{eqn:phase_original}. In this study, the Taylor series is truncated to a third-order Taylor polynomial, based on available information from the pulsar catalog provided by the ATNF \cite{ATNFPulsarCatalogue}.
\begin{equation}
\label{eqn:phase_original}
    \Phi_{\mb{b},t_0}(t_f)
        := \phi_{\mb{b},t_0}
         + \sum_{z=1}^\infty \dfrac{d^{(z-1)} f_{t_0}}{dt^{(z-1)}}
                     \cdot{} \dfrac{(t_f-t_0)^z}{z!}
\end{equation}

\subsubsection{Epoch Update.}
The timing model can be updated to center on a new epoch $t_1 = t_0 + \Delta{}t$. First, a new phase $\phi_{\mb{b},t_1}$ is computed using Eq.~\eqref{eqn:phase_update}. The second term on the right-hand side uses the floor function $\floor{x}$, which returns the largest integer that is smaller than $x$. This largest integer is used to reset the integer portion of the phase to zero, which may improve numerical stability. The Taylor series expansion centered on $t_1$ is then Eq.~\eqref{eqn:phase_time}, with frequency derivatives defined by Eq.~\eqref{eqn:frequency_update}.
\begin{equation}
\label{eqn:phase_update}
    \phi_{\mb{b},t_1} = \Phi_{\mb{b},t_0}(t_1) - \floor{\Phi_{\mb{b},t_0}(t_1)}
\end{equation}
\begin{equation}
\label{eqn:frequency_update}
\dfrac{d^n f_{t_1}}{dt^n}
    = \dfrac{d^n f_{t_0}}{dt^n} 
    + \sum_{z=1}^\infty \hspace{0.02in}\dfrac{d^{(n+z)} f_{t_0}}{dt^{(n+z)}} 
                 \cdot{}  \dfrac{(t_1-t_0)^z}{z!}
\end{equation}
\begin{equation}
\label{eqn:phase_time}
\Phi_{\mb{b},t_1}(t_f)
= \phi_{\mb{b},t_1}
+ \sum_{z=1}^\infty \dfrac{d^{(z-1)} f_{t_1}}{dt^{(z-1)}}
\cdot{} \dfrac{(t_f-t_1)^z}{z!}
\end{equation}

\subsubsection{Position Update.}
The timing model can also be re-centered about a new position $\mb{p}$. The difference in the time of arrival of a pulsar wavefront between an original position $\mb{b}$ and a new position $\mb{p}$ can be calculated from the Roemer and Shapiro delays. The wavefront time of arrival, $t_\mb{p}$, at the new position $\mb{p}$ is computed by the time transfer equation:
\begin{equation}
\label{eqn:time_transfer}
    t_\mb{p} = t_\mb{b}
             - \big(t_{\mb{b}} - t_\mb{p}\big)_{_{R\bm{_\odot}}}
             - \big(t_{\mb{b}} - t_\mb{p}\big)_{_{S\bm{_\odot}}}
\end{equation}

Assuming the clocks at $\mb{p}$ and $\mb{b}$ are synchronized and do not experience time dilation with respect to each other, the new Taylor series expansion can be expressed as Eq.~\eqref{eqn:phase_position}. While the new phase and frequency derivatives are based on $t_\mb{p}$, the epoch remains as $t_\mb{b}$. This is because the actual reference time did not change --- only the arrival pulse has changed due to the change in position. 
\begin{equation}
\label{eqn:phase_position}
    \Phi_{\mb{p},t_\mb{b}}(t_f)
    = \phi_{\mb{b},t_\mb{p}}
    + \sum_{z=1}^\infty \dfrac{d^{(z-1)} f_{t_\mb{p}}}{dt^{(z-1)}}
    \cdot{} \dfrac{(t_f-t_\mb{b})^z}{z!}
\end{equation}

\subsubsection{Epoch and Position Update.}
Finally, a combined change in both epoch and position may be achieved by changing the epoch to $t_1 = t_0 + \Delta{}t$, and updating the phase and frequency derivatives via $t^* = t_0 + \Delta{}t + \big(t_{\mb{b}} - t_\mb{p}\big)_{_{R\bm{_\odot}}} + \big(t_{\mb{b}} - t_\mb{p}\big)_{_{S\bm{_\odot}}}$:
\begin{equation}
\label{eqn:phase_both}
    \Phi_{\mb{p},t_1}(t_f)
    = \phi_{\mb{b},t^*} 
    + \sum_{z=1}^\infty \dfrac{d^{(z-1)} f_{t^*}}{dt^{(z-1)}}
    \cdot{} \dfrac{(t_f-t_1)^z}{z!}
\end{equation}
\section{Norm-Minimization Algorithm}
\label{sec:algorithm}

The proposed cold-start XNAV algorithm directly solves for candidate points that minimize the norm of distances to each observed pulsar wavefront. Doing so reduces the memory requirement associated with vertex tracking (as in Reference~\citenum{lohan2021methodology}) and simplifies accommodation of higher-fidelity models. For visual clarity, a two-dimensional example will be used to explain the algorithm.


\subsection{Wavefront Position}
To solve for candidate positions using observed pulsar wavefronts, the position of the wavefronts must first be determined. From observing pulsar signals, the fractional component of pulsar $i$'s observed phase, $\phi_i$, can be determined \cite{emadzadeh2011x,zhang2014new}. 

Let $d$ be the distance from the reference point to the m\textsuperscript{th} wavefront of the i\textsuperscript{th} pulsar. Let the observed phase of the i\textsuperscript{th} pulsar be $\phi_i$ at $t_\mb{p}$ in coordinate time (assuming that time dilation is taken into account since the onboard clock measures proper time), where $\mb{p}$ is the position of the spacecraft. Assume pulsar timing models are defined at $\mb{b}$ and $t_0$ in coordinate time. The spacecraft position, $\mb{p}$, which satisfies Eq.~\eqref{eqn:integer_phase_equality}, must be known for $d$ to be computed with $d = \hat{\mb{n}}\cdot{}\mb{p}$.
\begin{equation}
    \label{eqn:integer_phase_equality}
    m + \phi_i = \Phi_{\mb{b},t_0}(t_\mb{b})
\end{equation}

In Eq.~\eqref{eqn:integer_phase_equality}, $t_\mb{b}$ is computed from the time transfer equation shown in Eq.~\eqref{eqn:time_transfer}. Since $\mb{p}$ is unknown (by the nature of cold-start problems), $\mb{p}$ may be instead approximated using Eq.~\eqref{eqn:approx_p}, where the $\tilde{\mb{r}}$ is the approximation of relative position $\mb{r}$ as in Eq.~\eqref{eqn:approx_r}.
\begin{equation}
    \label{eqn:approx_p}
    \mb{p} \approx \tilde{\mb{p}} = \tilde{\mb{r}} + \mb{b}
\end{equation}
\begin{equation}
    \label{eqn:approx_r}
    \mb{r} \approx \tilde{\mb{r}} = \tilde{\mb{r}}_0 + \delta{r}\cdot{}\hat{\mb{n}}
\end{equation}

The initial guess, $\tilde{\mb{r}}_0$, is not strictly necessary but may improve accuracy; $\tilde{\mb{r}}_0$ can be obtained by applying L$_2$- or L$_\infty$-norm minimization as described in the following subsections. The objective is now to solve for the value of $\delta{r}$ that satisfies the equality shown in Eq.~\eqref{eqn:integer_phase_equality_approximation}:
\begin{equation}
    \label{eqn:integer_phase_equality_approximation}
    m + \phi_i = \Phi_{\mb{b},t_0}(\tilde{t_\mb{b}})
\end{equation}

Eq.~\eqref{eqn:integer_phase_equality_approximation} can be solved using an iterative scheme starting with $\delta{r}_0 = 0$, as shown in Eq.~\eqref{eqn:get_distance}. Note that $\tilde{t_\mb{b}}_{,j}$ is an approximation of $t_\mb{b}$ --- it is computed using Eq.~\eqref{eqn:time_transfer} with $\tilde{\mb{p}}_j$ at each iteration, which updates the estimates of the Roemer and Shapiro delays. In other words, Eq.~\eqref{eqn:get_distance} allows the norm-minimization algorithm to account for the Roemer delay, Shapiro delay, and higher-order pulsar timing models.
\begin{equation}
    \label{eqn:get_distance}
    \begin{aligned}
        \tilde{\mb{p}}_j &=\tilde{\mb{r}}_0+\delta{r}_j\cdot{}\hat{\mb{n}}+\mb{b}\\
        \tilde{\phi}_{i,j} &= \Phi_{\mb{b},t_0}(\tilde{t_\mb{b}}_{,j}) \\
        \delta{r}_{j+1} &= \delta{r}_j + \Big[(m+\phi_i)-\tilde{\phi}_{i,j}\Big]\cdot{}\dfrac{c}{f_{t_0}}
    \end{aligned}
\end{equation}

Although it is not self-evident that the iterative process converges to some value of $\delta{r}$, this is indeed the case. First, we assert that $\Phi_{\mb{b},t_0}(\tilde{t_\mb{b}})$ is a monotonically increasing function for small variations in $\tilde{t_\mb{b}}$ and an appropriately chosen $t_0$. This is based on the fact that $\Phi_{\mb{b},t_0}$ is a Taylor series approximation of the pulsar phase around time $t_0$, and the pulsar phase always increases monotonically with time. We assume that the Taylor series is appropriately centered to accurately compute phases near $t_\mb{b}$.

Next, we assert that the time transfer equation, Eq.~\eqref{eqn:time_transfer}, forms a monotonic relation between $\tilde{t_\mb{b}}$ and $\delta{r}$. The time transfer equation, in the present notation, is:
\begin{equation}
    \tilde{t_\mb{b}} = t_{\mb{p}}
    + \big(\tilde{t_\mb{b}} - t_{\tilde{\mb{p}}}\big)_{_{R\bm{_\odot}}}
    + \big(\tilde{t_\mb{b}} - t_{\tilde{\mb{p}}}\big)_{_{S\bm{_\odot}}}
\end{equation}

Eqs.~\eqref{eqn:roemer_est}-\eqref{eqn:shapiro_est} show how the approximation $\tilde{\mb{r}} = \tilde{\mb{r}}_0 + \delta{r}\cdot{}\hat{\mb{n}}$ impacts the estimates of the Roemer and Shapiro delays. Note that in Eq.~\eqref{eqn:shapiro_est}, $\tilde{\mb{p}}_0 = \mb{b} + \tilde{\mb{r}}_0$.
\begin{equation}
    \label{eqn:roemer_est}
    \begin{aligned}
        \big(\tilde{t_\mb{b}} - t_{\tilde{\mb{p}}}\big)_{_{R\bm{_\odot}}}
        &=
        \dfrac{\hat{\mb{n}}\cdot{}\tilde{\mb{r}}_0}{c} + \dfrac{\delta{r}}{c} \\
        &+
        \dfrac{1}{2cD}
        \bigg\{
        &&\hspace{-0.4in}\+
        \Big[
        (\hat{\mb{n}}\cdot{}\tilde{\mb{r}}_0)^2
        + 2\delta{r}(\hat{\mb{n}}\cdot{}\tilde{\mb{r}})
        + \delta{r}^2
        \Big]
        -
        \Big[
        \tilde{r}_0^2
        + \delta{r}^2
        + 2\delta{r}(\hat{\mb{n}}\cdot{}\tilde{\mb{r}})
        \Big] \\
        &&&\hspace{-0.49in}+
        \Big[
        2(\hat{\mb{n}}\cdot{}\mb{b})(\hat{\mb{n}}\cdot{}\tilde{\mb{r}}_0)
        + 2(\hat{\mb{n}}\cdot{}\mb{b})\delta{r}
        \Big]
        -
        \Big[
        2(\mb{b}\cdot{}\tilde{\mb{r}}_0)
        + 2(\hat{\mb{n}}\cdot{}\mb{b})\delta{r}
        \Big]
        \bigg\}
    \end{aligned}
\end{equation}

\begin{equation}
    \label{eqn:shapiro_est}
    \big(\tilde{t_\mb{b}} - t_{\tilde{\mb{p}}}\big)_{_{S\bm{_\odot}}}
    =
    \dfrac{2\mu}{c^3}
    \ln{\Bigg\lvert
        \dfrac{ (\hat{\mb{n}}\cdot{}\tilde{\mb{p}}_0 + \delta{r}) + 
            \sqrt{ \vphantom{0_0^{0}}
                \tilde{p}_0^2
                +\delta{r}^2
                +2(\hat{\mb{n}}\cdot{}\tilde{\mb{p}}_0)\delta{r}}
        }
        {\hat{\mb{n}}\cdot\mb{b}+b}
        \Bigg\rvert}
\end{equation}

Taking the partial derivative of the Roemer delay with respect to $\delta{r}$ yields:
\begin{equation}
    \dfrac{d}{d\delta{r}}
    \big(\tilde{t_\mb{b}} - t_{\tilde{\mb{p}}}\big)_{_{R\bm{_\odot}}}
    = \dfrac{1}{c} > 0
\end{equation}

For Shapiro delay, since the natural log is a monotonically increasing function and $\hat{\mb{n}}\cdot{}\mb{b}+b \geq 0$, it is sufficient to examine the numerator to understand the behavior of the Shapiro delay. If the denominator were zero, it would mean that the reference point is in opposition with the observed pulsar (with respect to the gravitational source), in which case the pulsar cannot be observed. Furthermore, recall that the numerator is an expansion of $\hat{\mb{n}}\cdot{}\tilde{\mb{p}}+\tilde{p}$ (where $\tilde{\mb{p}} = \tilde{\mb{p}}_0+\delta{r}\cdot{}\hat{\mb{n}}$), which is also non-negative. Therefore the absolute value can be discarded without affecting the analysis. Continuing, let the numerator in the natural logarithm of Eq.~\eqref{eqn:shapiro_est} be $g(\delta{r})$:
\begin{equation}
    \dfrac{d}{d\delta{r}}g(\delta{r})
    = 1 + \dfrac{\delta{r}+\hat{\mb{n}}\cdot{}\tilde{\mb{p}}_0}
                {\sqrt{ \vphantom{0_0^{0}}
                        \tilde{p}_0^2
                        +\delta{r}^2
                        +2(\hat{\mb{n}}\cdot{}\tilde{\mb{p}}_0)\delta{r}}} \geq 0
\end{equation}

\begin{figure}[!h]
    \centering
    \includegraphics[width=0.6\linewidth,
                     clip=true,
                     trim={200 40 200 70}]
                    {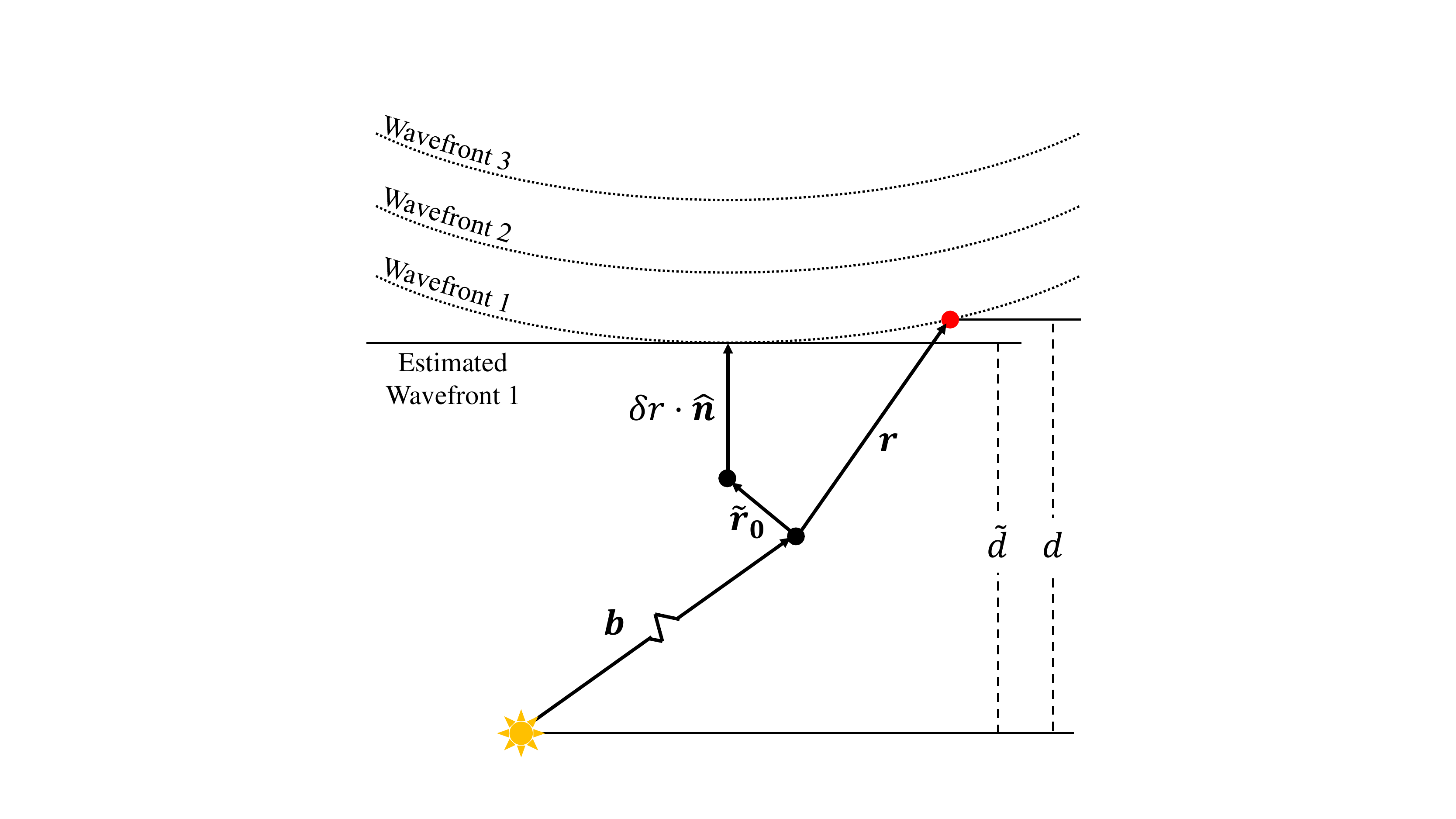}
    \caption{The error between the true spacecraft position $\mb{r}$ and approximated position $\tilde{r}\hat{\mb{n}}$ is reflected in errors in wavefront distance estimate $\tilde{d}$.}
    \label{fig:wavefront_estimate}
\end{figure}

We now conclude that $\tilde{t_\mb{b}}$ is strictly monotonically increasing with respect to $\delta{r}$. In turn, $\Phi_{\mb{b},t_0}$ monotonically increases with respect to $\delta{r}$. This means that a line search can be performed to find a $\delta{r}$ that produces the observed phase $(m+\phi_i)$. The iterative approach in Eq.~\eqref{eqn:get_distance} is one such implementation. Figure~\ref{fig:wavefront_estimate} shows the difference between the true wavefront position $d$ and the estimated position $\tilde{d}$ using the above procedure.

The approximation $\tilde{\mb{r}} = \tilde{\mb{r}}_0 + \delta{r}\cdot{}\hat{\mb{n}}$ becomes less accurate as $r$ increases, that is, for wavefronts farther from the reference point $\mb{b}$. This degradation in the positioning of pulsar wavefronts can cause the algorithm to fail. A later section will examine the maximum value of $r$ that will produce correct solutions.

\FloatBarrier
\subsection{L$_\infty$-norm Minimization}

The objective of XNAV is to find a candidate point that falls within the banded-error bound of all observed pulsars. This can be achieved through L$_\infty$ minimization: find a candidate point that minimizes the maximum quotient of its distance to each pulsar's nearest wavefront divided by the width of the pulsar's error band. Restated mathematically:

Let unit vectors ${\hat{\mb{n}}_1, \hat{\mb{n}}_2, \cdots, \hat{\mb{n}}_{_N}}$ point toward pulsars 1 to $N$. Again, we assume that pulsars are sufficiently distant such that the unit vector $\hat{\mb{n}}$ pointing toward the pulsar is unchanged within the Solar System. Let $\epsilon_1, \epsilon_2, \cdots, \epsilon_{_N}$ be the half-width of the banded-error bounds for each pulsar, and let $\tilde{d}_1, \tilde{d}_2, \cdots, \tilde{d}_{_N}$ be the estimated distances from the reference point to one wavefront of each pulsar. Therefore, we can state that the candidate position must satisfy the following optimization problem:
\begin{equation}
    \begin{alignedat}{2}
        \mathrm{minimize}   \quad & s \\
        \mathrm{s.t.}       \quad
        & \abs{\hat{\mb{n}}_i^T \tilde{\mb{x}} - \tilde{d}_i} \:&& \leq s\cdot{}\epsilon_1
        \quad\forall\; i \in [1,N]
    \end{alignedat}
\end{equation}
where $\tilde{\mb{x}}$ is the candidate point. The problem may be solved using the dual simplex method after the following reformulation:
\begin{equation}
\label{eqn:dual_simplex}
    \begin{alignedat}{2}
        \mathrm{minimize}   \quad & s \\
        \mathrm{s.t.}       \quad
        & \+ \hat{\mb{n}}_i^T   \tilde{\mb{x}} - \tilde{d}_i    &&\leq s\cdot{}\epsilon_1
        \quad\forall\; i \in [1,N] \\
        & \- \hat{\mb{n}}_i^T   \tilde{\mb{x}} + \tilde{d}_i    &&\leq s\cdot{}\epsilon_1
        \quad\forall\; i \in [1,N]
    \end{alignedat}
\end{equation}

If $s > 1$ in the solution to Eq.~\eqref{eqn:dual_simplex}, there is no value of $\tilde{\mb{x}}$ that can satisfy the banded-error bounds of all pulsars simultaneously. Conversely, this means a value of $\tilde{\mb{x}}$ that satisfies $s \leq 1$ will always be found if it exists.

\subsection{L$_2$-norm Minimization}

The dual simplex method is computationally expensive compared to techniques for solving other linear systems, such as linear least squares. Since the search space for XNAV in a cold-start scenario may span many cubic AU, a volume with a major axis several orders of magnitude greater than the wavelength of each pulsar, the total number of wavefront combinations is potentially large and warrants consideration of computational efficiency. An alternative method that quickly filters out most combinations can significantly reduce computational resource requirements.

Rather than minimizing the L$_\infty$-norm, the L$_2$-norm may be minimized instead. In other words, a candidate point can be found from solving $\mb{A}\tilde{\mb{x}}=\mb{y}$ via linear least squares, with:
\begin{equation}
    \begin{aligned}
        \mb{A} &= [\hat{\mb{n}}_1,\hat{\mb{n}}_2,\cdots,\hat{\mb{n}}_{_N}]^T \\
        \mb{y} &= [\tilde{d}_1,\tilde{d}_2,\cdots,\tilde{d}_{_N}]^T
    \end{aligned}
\end{equation}

Linear least-squares produces a different candidate point from that produced by minimizing the L$_\infty$-norm. As a simple example, consider the following system with solutions that minimize ${\norm{\mb{A}\tilde{\mb{x}}-\mb{y}}_2}$ and $\norm{\mb{A}\tilde{\mb{x}}-\mb{y}}_\infty$, respectively:
\begin{equation}
    \mb{A} = 
    \left[\begin{array}{cc}
        \+1 & 0 \\
        \+0 & 1 \\
        \+1 & 1 \\
        \-1 & 1
    \end{array}\right]
    \;,\;
    \mb{y} = 
    \left[\begin{array}{c}
        0 \\ 0 \\ 3 \\ 3
    \end{array}\right]
    \quad\Rightarrow\quad
    \tilde{\mb{x}}_{_{L_2}} = 
    \left[\begin{array}{c}
    0 \\ 2
    \end{array}\right]
    \;,\;
    \tilde{\mb{x}}_{_{L_\infty}} = 
    \left[\begin{array}{c}
    0 \\ 1.5
    \end{array}\right]
\end{equation}

The largest element in the residual of the L$_2$-norm is 2, while the largest element in the residual of the L$_\infty$-norm is 1.5. The L$_\infty$-norm minimization method found a point whose greatest distance to the pulsar wavefronts was smaller than that found by the linear least squares approach. It is therefore possible for the linear least squares approach to miss candidate points that would have been found by L$_\infty$-norm minimization. In the implementation of the L$_2$-norm minimization within the algorithm, the widths of the error bands are scaled by a factor of 5 to mitigate this shortcoming. 

Despite its inferior accuracy, linear least-squares is valuable due to its speed. Since the matrix $\mb{A}$ is comprised of row vectors pointing toward each pulsar, linear least squares can be accelerated by pre-computing the QR decomposition of $\mb{A}$ as follows:
\begin{equation}
    \mb{A} = \mb{QR} 
    = \mb{Q} \left[\begin{array}{c} \mb{R}_1 \\ \mb{0} \end{array}\right]
    = [\mb{Q}_1 \; \mb{Q}_2] \left[\begin{array}{c} \mb{R}_1 \\ \mb{0} \end{array}\right]
    = \mb{Q}_1 \mb{R}_1
\end{equation}
where $\mb{Q}_1$ is an orthogonal matrix and $\mb{R}_1$ is an upper triangular matrix. This applies to an $m$-by-$n$ matrix $\mb{A}$ with $m \geq n$. The precomputed $\mb{Q}_1$ and $\mb{R}_1$ matrices can then be used to rapidly solve the original problem through back-substitution after rewriting it as the following:
\begin{equation}
\label{eqn:backsub}
    \begin{aligned}
        \mb{A}\tilde{\mb{x}} &= \mb{y} \\
        \mb{Q}_1\mb{R}_1\tilde{\mb{x}} &= \mb{y} \\
        \mb{R}_1\tilde{\mb{x}} &= \mb{Q}_1^T\mb{y}
    \end{aligned}
\end{equation}

\subsection{Recursive Iteration}

\begin{figure}[ht!]
    \centering
    \includegraphics[width=\linewidth,
                     clip=true,
                     trim={70 50 70 50}]
                    {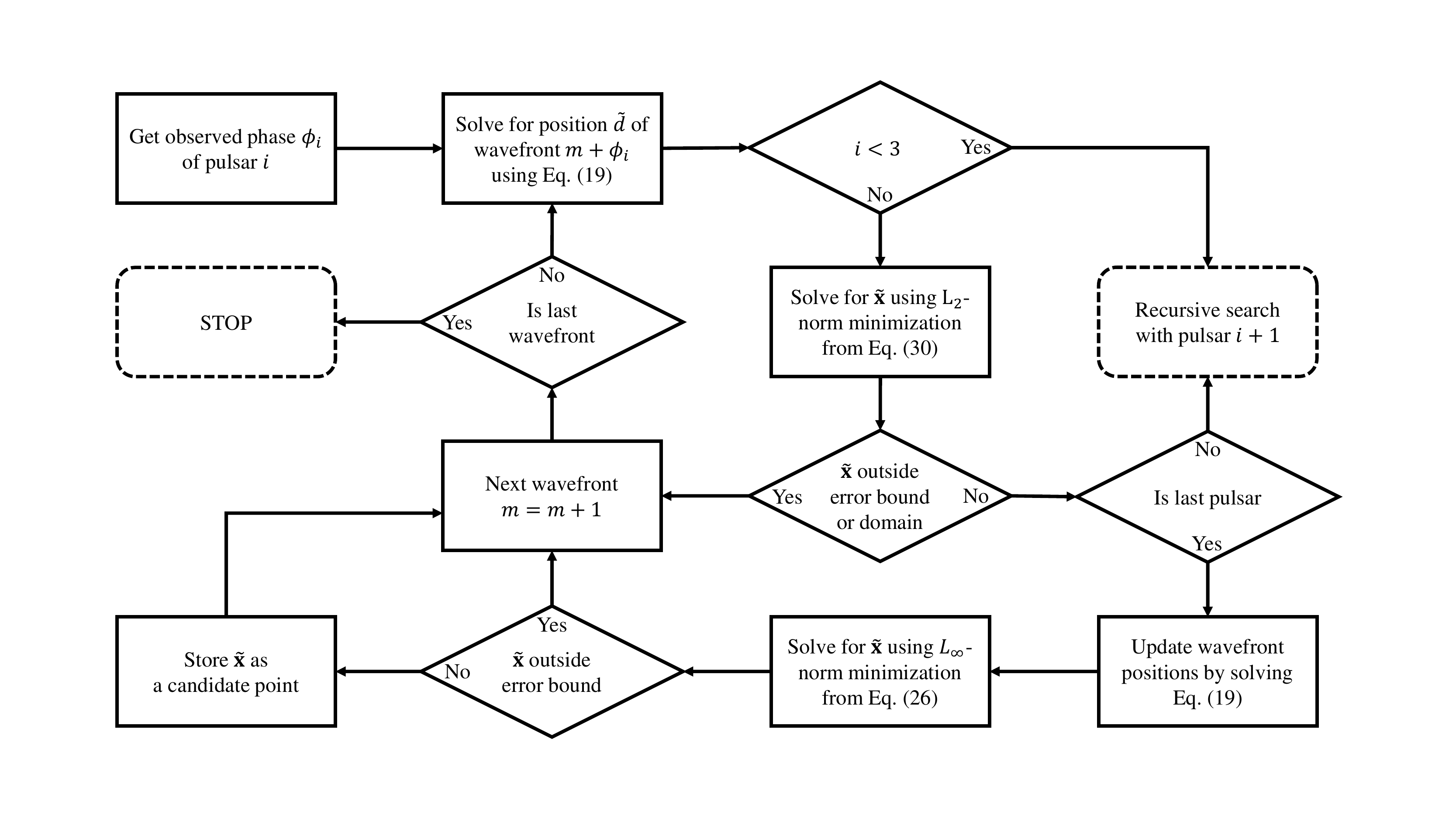}
    \setlength{\belowcaptionskip}{-5pt}
    \caption{Recursive wavefront search algorithm without pivoting.}
    \label{fig:algo_naive}
\end{figure}

To find all candidate points within a spatial domain, all possible combinations of pulsar wavefronts that pass through the domain must be explored, as shown in Figure~\ref{fig:algo_naive}. The algorithm depicted is recursive --- it iterates through all wavefronts from one pulsar that are within the spatial domain. For each wavefront, the algorithm calls itself to iterate through the wavefronts of the next pulsar. 

For the $i$\textsuperscript{th} pulsar during recursion, only the vectors pointing to the first $i$ pulsars are included for QR decomposition. Thus, the QR decomposition of every matrix formed by the first $i$ ($\textrm{dim} \leq i \leq N$) rows of $\mb{A}$ must be pre-computed and stored, where dim is the dimensionality of the problem, i.e. either 2 or 3, corresponding to two-dimensional space and three-dimensional space, respectively, and $N$ is the total number of observed pulsars. For example, for a three-dimensional problem:
\begin{equation}
\begin{aligned}
    \mb{Q}_3\mb{R}_3 &= [\hat{\mb{n}}_1,\hat{\mb{n}}_2,\hat{\mb{n}}_3]^T \\
    & \vdots \\
    \mb{Q}_{_N}\mb{R}_{_N} &= [\hat{\mb{n}}_1,\hat{\mb{n}}_2,\cdots,\hat{\mb{n}}_{_N}]^T = \mb{A}
\end{aligned}
\end{equation}

\subsection{Wavefront Pivoting}

\begin{figure}[ht!]
    \centering
    \includegraphics[width=0.65\linewidth]{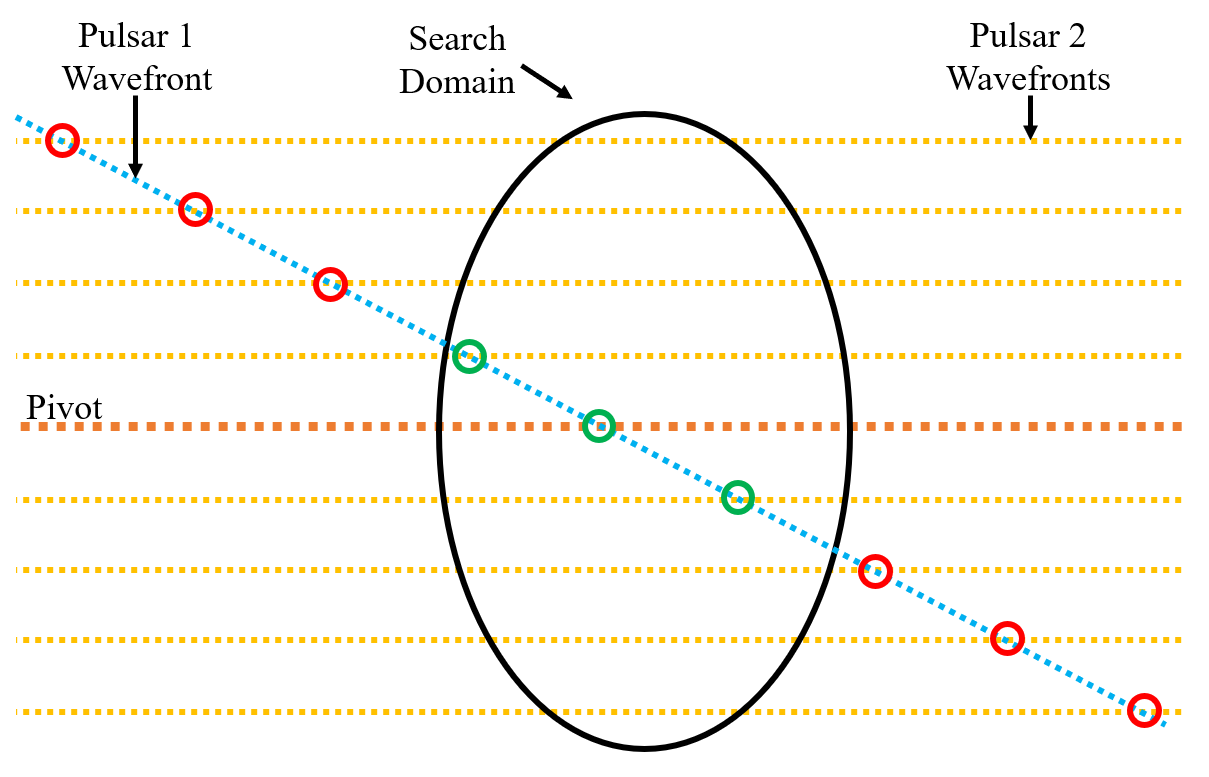}
    \setlength{\belowcaptionskip}{-5pt}
    \caption{Intersecting a wavefront from pulsar 1 (blue) with wavefronts of pulsar 2 (yellow) creates a collection of intersection points whose majority lies outside the search domain (black). }
    \label{fig:pivot}
\end{figure}

While iterating through all wavefront combinations is reliable, a large number of intersections may be outside the domain. For example, in Figure~\ref{fig:pivot}, red circles represent intersection points outside the domain and green circles represent intersections inside the domain. While most intersections are outside the domain, all intersections must be considered using the method shown in Figure~\ref{fig:algo_naive}. The time spent calculating the linear least-squares solution for intersections outside the domain is unnecessary.

Instead, a pivot point can be selected, which is the wavefront that would generate a point closest to the center of the domain, as shown in Figure~\ref{fig:pivot}. If the domain is convex, then the algorithm only has to search until a point is outside the domain, because all the remaining intersections are guaranteed to be outside the domain. This technique can greatly reduce the number of unnecessary computations for arbitrary convex search domain shapes.

\begin{figure}[ht!]
    \centering
    \includegraphics[width=\linewidth,
                     clip=true,
                     trim={70 50 70 50}]
                    {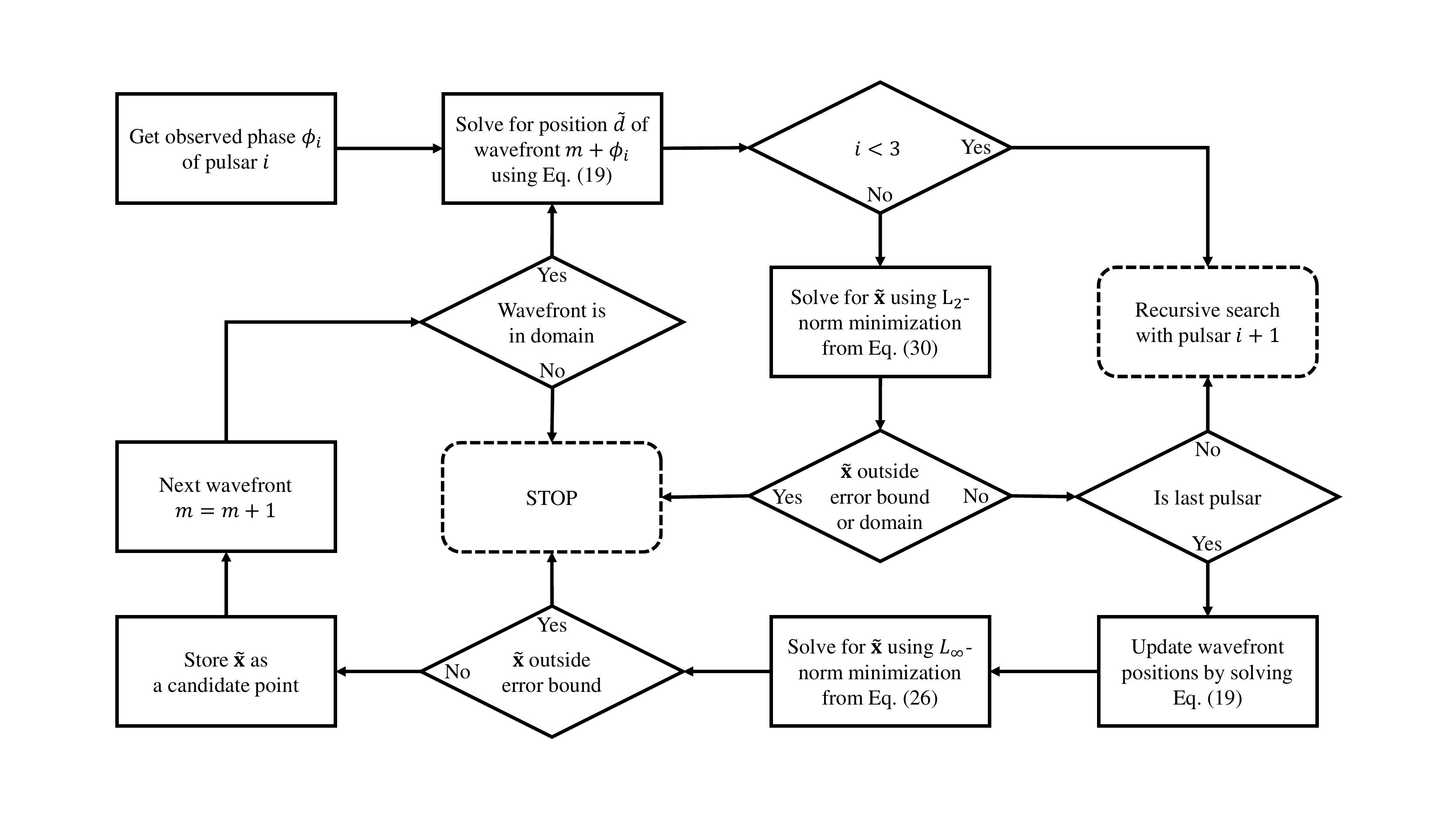}
    \setlength{\belowcaptionskip}{-5pt}
    \caption{Recursive wavefront search algorithm with pivoting.}
    \label{fig:algo_pivoted}
\end{figure}

If the recursive algorithm has found a point $\mb{x}$ that satisfies the banded-error bounds, the pivot point is instead chosen to be closest to $\mb{x}$ for the next layer of recursion. Once again, the algorithm only needs to search until a point fails to satisfy the error bounds, since all remaining points would be even farther away. With the new choice of pivot point, the recursive algorithm (see Figure~\ref{fig:algo_naive}) can be modified as shown in Figure~\ref{fig:algo_pivoted}. Rather than continuing to the next wavefront when an intersection point is outside the error bound or domain, the algorithm can simply terminate the current recursive layer, and return to the previous layer.

\FloatBarrier
\section{Methodology}
\label{sec:methodology}

The remainder of this paper will analyze how the reliability and accuracy of the norm-minimization XNAV algorithm are impacted by higher-fidelity models and observation options. This section describes the baseline trajectory and pulsars used for Monte Carlo simulations.

\subsection{Reference Frames}
The reference frame used is the International Celestial Reference Frame (ICRF), also commonly referred to as the J2000 frame, which matches that used by the Jet Propulsion Laboratory's SPICE toolkit \cite{SPICECoordOverview}. The ICRF is centered at the SSB. The reference time uses Modified Julian Dates (MJD) with Barycentric Dynamical Time (TDB) to be compatible with the pulsar timing models obtained from the ATNF \cite{GSFCTimingSystems,ATNFPulsarCatalogue}.

\subsection{Simulation Framework}

\subsubsection{Truth Model.}
The spacecraft initial state is defined and propagated using the General Mission Analysis Tool (GMAT)\cite{gmat}. The Einstein delay is computed from the trajectory of the spacecraft generated by GMAT. Next, pulsar observations are simulated from the spacecraft's terminal position. Currently, the simulation assumes instantaneous, concurrent observations of multiple pulsars. The observed phase of each pulsar is computed using Eq.~\eqref{eqn:phase_both}. Higher-fidelity pulsar timing models including the Roemer delay, parallax effect, and higher-order pulsar frequency model are always included in the truth model. 

\subsubsection{Perturbations.}
Perturbations to the truth model arise from inaccuracies in the observed phase of a pulsar signal. Observation noise is sampled from independent zero-mean Gaussian distributions with a standard deviation of $1.0\times10^{-3}$ that is added to the true observed phase of each pulsar. The perturbed phase observations are supplied to the norm-minimization algorithm to solve for candidate spacecraft positions. 

\subsubsection{Algorithm Model.}
The norm-minimization algorithm accounts for higher-fidelity pulsar timing models as described by Eq.~\eqref{eqn:get_distance}. In addition, each higher-fidelity model can be selectively disabled to determine whether the exclusion of certain models hinders the algorithm's ability to identify correct candidate positions. Disabling a certain model means setting its contribution toward the pulsar timing model to zero. For example, disabling the Shapiro delay means letting $\Delta{}S\bm{_\odot} = 0$.

\subsubsection{Monte Carlo Analysis}

Uncertainty analysis is conducted using Monte Carlo techniques. The sample size of Monte Carlo simulations was chosen to be 300, because the candidate position statistics do not vary significantly when the sample size is further increased.

\subsection{Orbit and Pulsar Selection}

\begin{figure}[!h]
    \centering
    \setlength\belowcaptionskip{-5pt}
    \includegraphics[width=0.65\linewidth,
    clip=true,
    trim={12 16 30 40}]
    {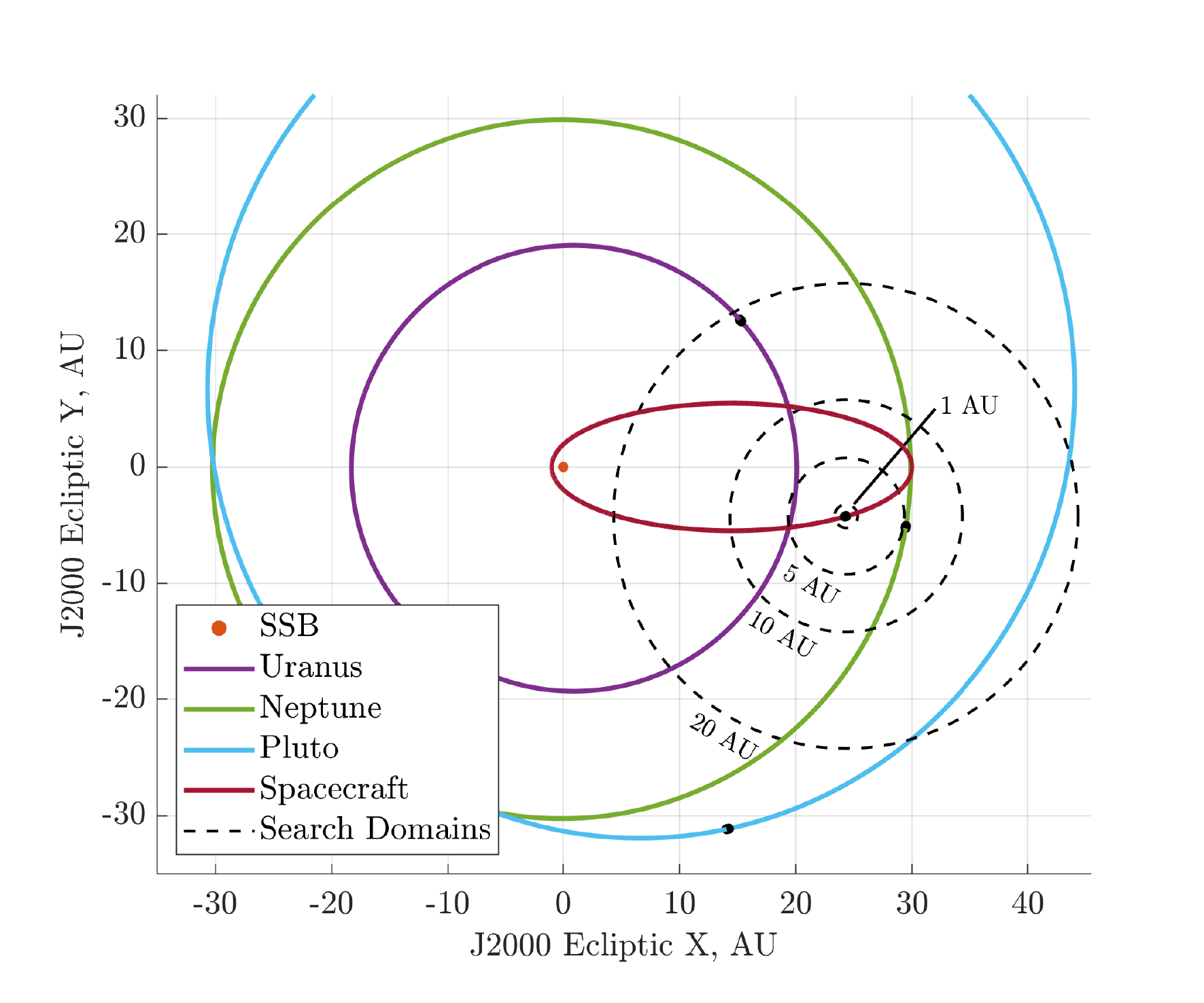}
    \caption{The simulated positions of the spacecraft and nearby celestial bodies are shown in the J2000 Ecliptic coordinate system at MJD 59215.5 with their respective trajectories.}
    \label{fig:simulated_orbit}
\end{figure}

The orbit chosen for analysis is an Earth-Neptune transfer trajectory with the spacecraft at true anomaly $f=\ang{170}$. In the J2000 frame, the terminal position of the spacecraft is $[24.332, -3.861,\allowbreak -1.719]$ AU with velocity $[3.656,0.963,0.429]$ \si{\kilo\meter\per\second}. The trajectory of the spacecraft is shown in Figure~\ref{fig:simulated_orbit}. The ellipsoidal search domain, centered at the position of the spacecraft, is shown with varying semi-major axis radii of 1, 5, 10, and 20 AU. The spacecraft onboard clock's latest synchronization is simulated to occur 60 days prior to pulsar observations. This means that the spacecraft trajectory has been unknown for the past 60 days, so proper time and coordinate time have diverged due to time dilation.

Two sets of nine pulsars are used to evaluate the performance of the algorithm, as shown in Table~\ref{tab:pulsars}: a low-frequency pulsar set and a mixed-frequency pulsar set. These pulsars were chosen from a list of 34 pulsars used in previous feasibility studies \cite{shemar2016towards,lohan2021methodology}. The low-frequency set includes four pulsars that minimized the number of candidate points as discussed by Lohan in Reference~\citenum{lohan2021methodology}. The mixed-frequency set includes three high frequency pulsars previously used by SEXTANT \cite{winternitz2016sextant}. The right ascension in the J2000 frame (RAJ) is denoted in hour-minute-second (hms) format, while the declination in the J2000 frame (DECJ) is denoted in degree-minute-second (dms) format. $f$ is the first-order coefficient of the pulsar phase Taylor polynomial as in $\frac{d^{(0)}f_{t^*}}{dt^{(0)}}$ in Eq.~\eqref{eqn:phase_both}.

\begin{table}[!h]
\centering
\caption{Key characteristics of pulsars that compose the low- and mixed-frequency pulsar sets \cite{ATNFPulsarCatalogue}.}
\label{tab:pulsars}
\begin{tabular}{|l|r|r|r|c|c|}
\hline
Pulsar Name & RAJ (hms)   & DECJ (dms)   & $f$ (Hz)& Pulsar Set & Note\\ \hline
J1119-6127  & 11:19:14.30 & -61:27:49.50 & 2.451   & Both       & -\\ \hline
J1846-0258  & 18:46:24.94 & -02:58:30.10 & 3.062   & Both       & Lohan\\ \hline
J0631+1036  & 06:31:27.52 & +10:37:02.50 & 3.475   & Both       & Lohan\\ \hline
J0633+1746  & 06:33:54.15 & +17:46:12.91 & 4.218   & Both       & Lohan\\ \hline
B1929+10    & 19:32:14.06 & +10:59:33.38 & 4.415   & Both       & -\\ \hline
J1930+1852  & 19:30:30.13 & +18:52:14.10 & 7.307   & Both       & -\\ \hline
J1811-1925  & 18:11:29.20 & -19:25:28.00 & 15.464  & Low        & Lohan\\ \hline
J2229+6114  & 22:29:05.28 & +61:14:09.30 & 19.371  & Low        & -\\ \hline
B0540-69    & 05:40:11.20 & -69:19:54.17 & 19.775  & Low        & -\\ \hline
B1821-24A   & 18:24:32.01 & -24:52:10.88 & 327.406 & Mixed      & SEXTANT\\ \hline
J0437-4715  & 04:37:15.90 & -47:15:09.11 & 173.688 & Mixed      & SEXTANT\\ \hline
B1937+21    & 19:39:38.56 & +21:34:59.12 & 641.928 & Mixed      & SEXTANT\\ \hline
\end{tabular}
\end{table}

\subsection{Higher Fidelity Models and Domain}
Unless otherwise mentioned, all higher-fidelity models discussed in Eqs.~\eqref{eqn:doppler_delay}-\eqref{eqn:phase_both} are included in the algorithm. Time dilation is estimated with a fixed error of 10 \si{\us}. The spatial search domain is an oblate spheroid whose minor axis is 1/1000th the length of its major axis. The search domain is very flat in one dimension since most orbits are in or near the ecliptic plane. The semi-major axis of the spheroid domain is nominally set to 1 AU. Other convex domain shapes are also possible, e.g., spheres.


\FloatBarrier
\section{Algorithm Performance}
\label{sec:performance}

Using the chosen trajectory and pulsars, 300-sample Monte Carlo simulations are performed to assess the performance of the algorithm given varying measurement qualities, algorithm models, and pulsar sets. Algorithm performance may be characterized by two metrics --- the likelihood that a unique, correct candidate point is found and the distance between the true position and the candidate point. A unique candidate means that the norm-minimization algorithm finds a single combination of wavefronts that produces a candidate point within all error bands within the search domain. A correct candidate means that the candidate point lies within the same set of error bands as the true position.

\subsection{Phase Uncertainty and Pulsar Selection}

Errors in the observed phase $\tilde{\phi}$ of a pulsar wavefront lead to errors in the positioning of pulsar wavefront $\tilde{d}$, which in turn may cause inaccuracies in the norm-minimization algorithm. The wavefront position error can be approximated to first order by Eq.~\eqref{eqn:wavefront_error}, where $c$ is the speed of light and $f$ is the pulsar frequency. 
\begin{equation}
    \label{eqn:wavefront_error}
    \abs{d-\tilde{d}} = \abs{\phi-\tilde{\phi}}\cdot{}(c/f)
\end{equation}

\begin{figure}[h!]
    \centering
    \begin{subfigure}{0.48\textwidth}
        \centering
        \includegraphics[width=\linewidth]{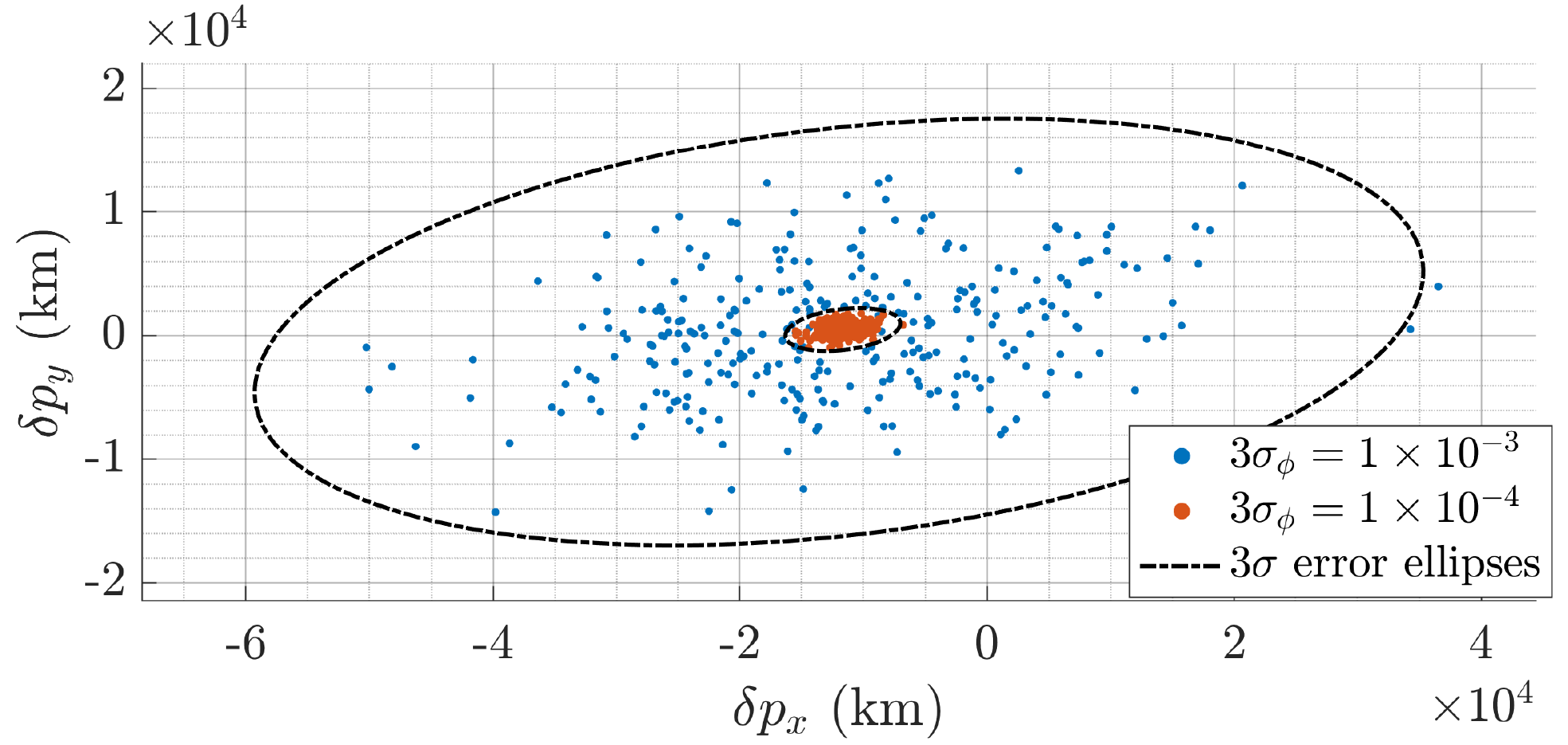}
        \caption{Low-Frequency Set, XY View}
    \end{subfigure}
    \hfill
    \begin{subfigure}{0.48\textwidth}
        \centering
        \includegraphics[width=\linewidth]{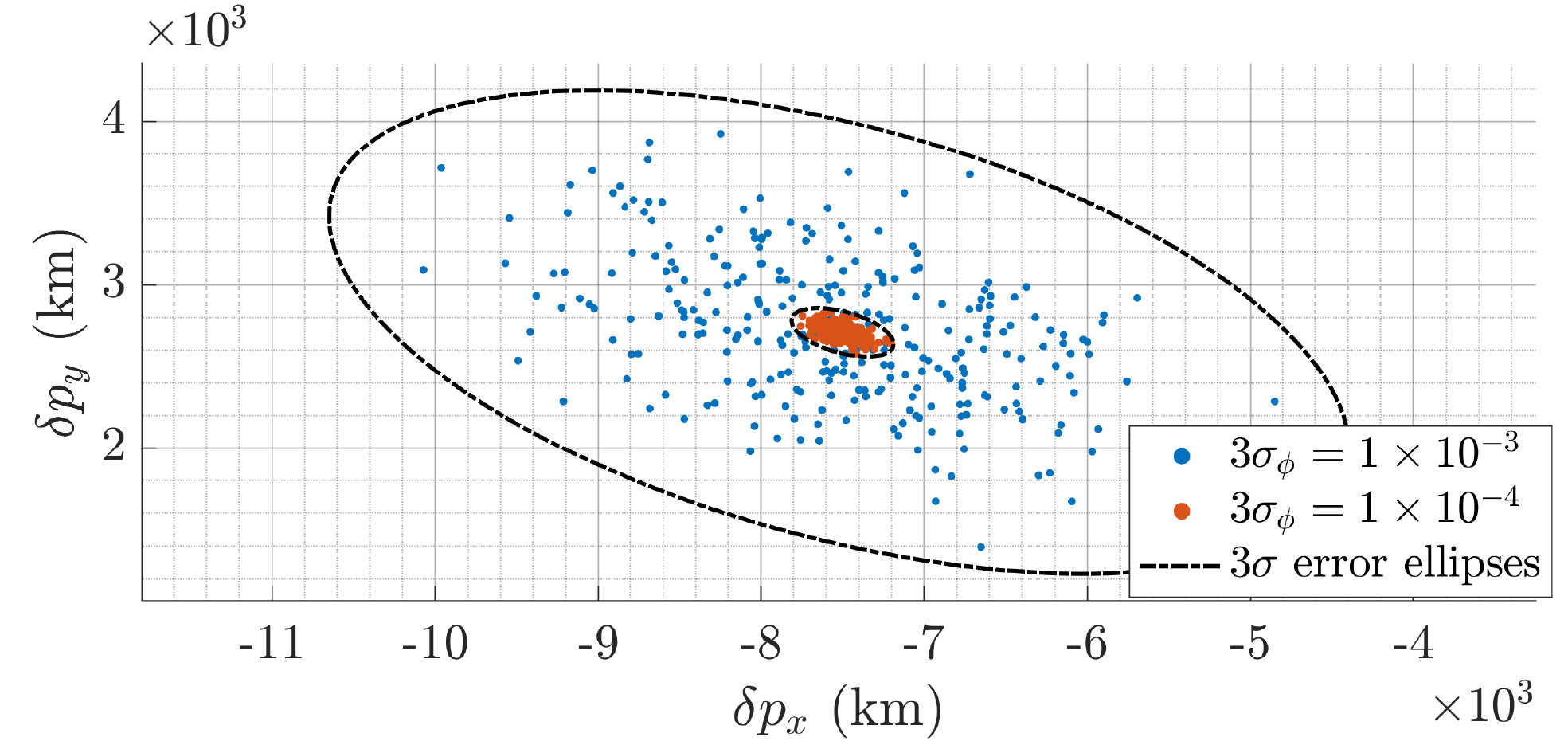}
        \caption{Mixed-Frequency Set, XY View}
    \end{subfigure}
    \vskip\parskip
    \begin{subfigure}{0.48\textwidth}
        \centering
        \includegraphics[width=\linewidth]{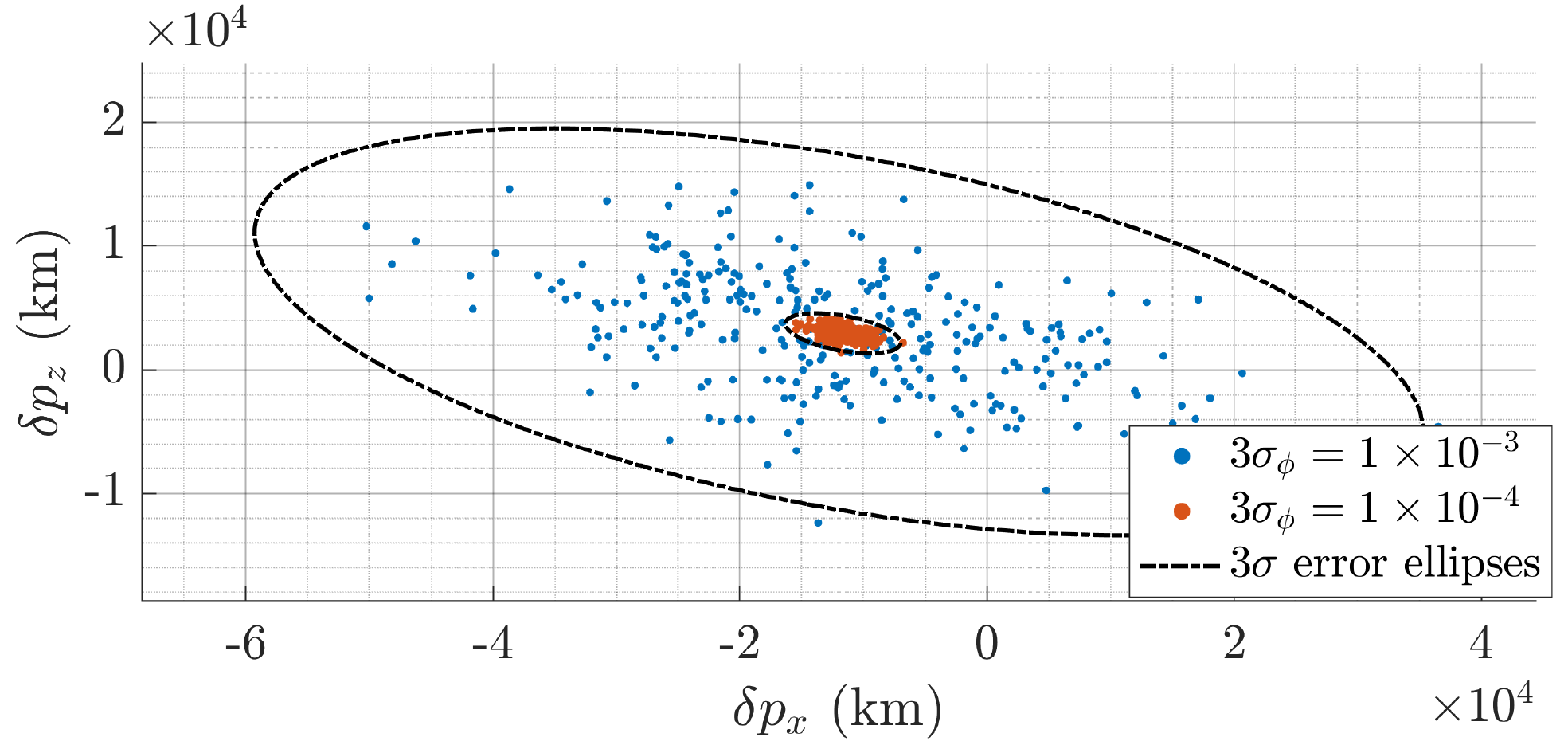}
        \caption{Low-Frequency Set, XZ View}
    \end{subfigure}
    \hfill
    \begin{subfigure}{0.48\textwidth}
        \centering
        \includegraphics[width=\linewidth]{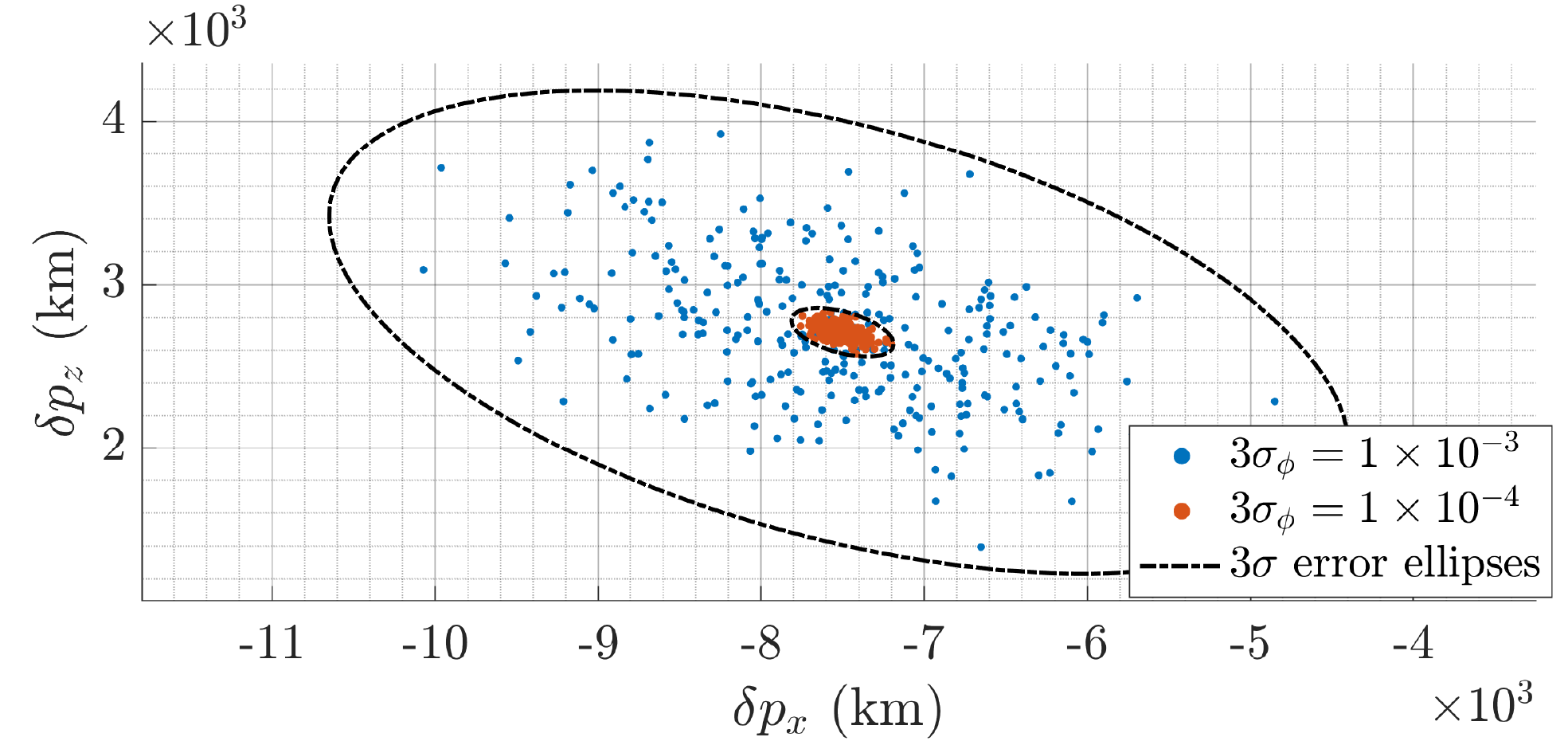}
        \caption{Mixed-Frequency Set, XZ View}
    \end{subfigure}
    \vskip\parskip
    \begin{subfigure}{0.48\textwidth}
        \centering
        \includegraphics[width=\linewidth]{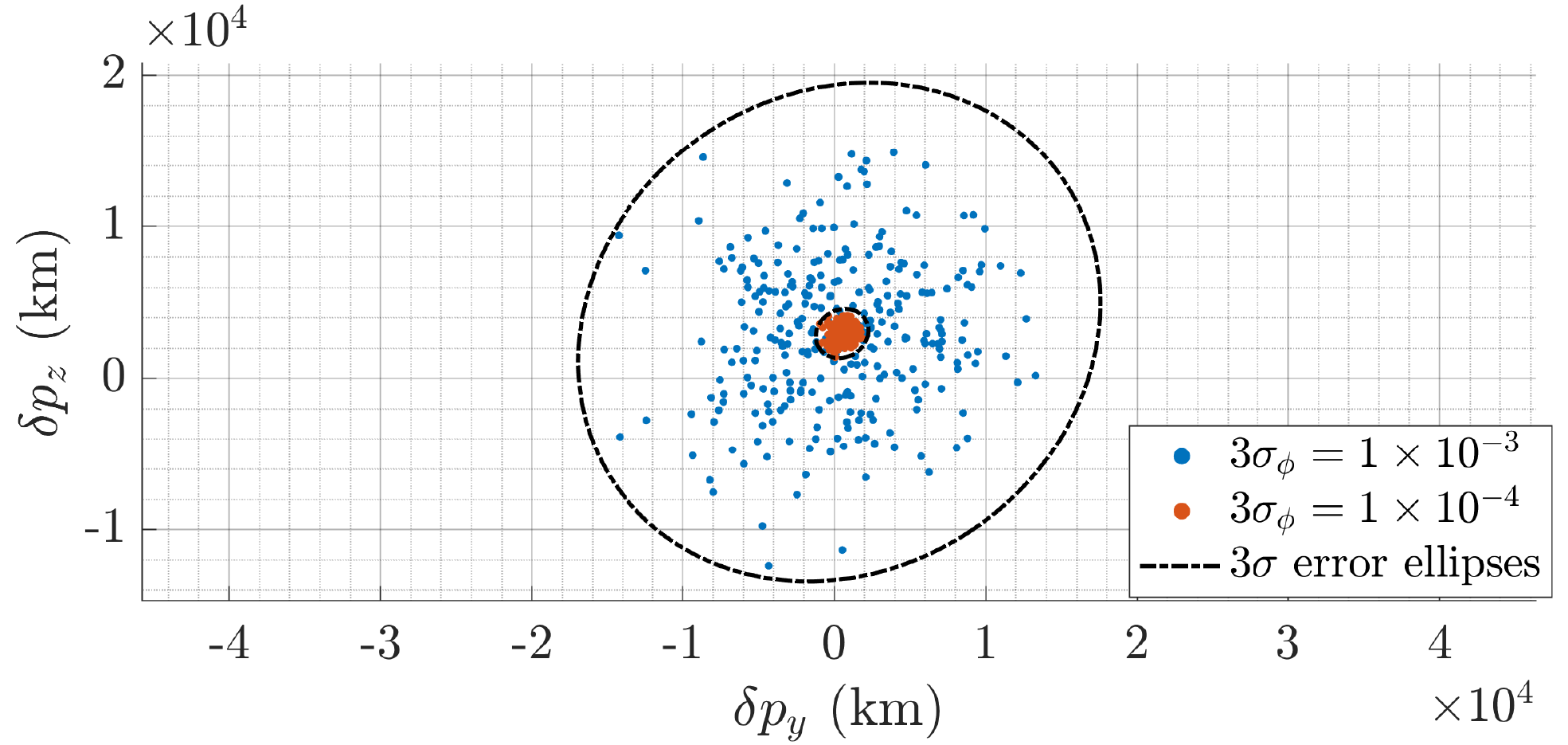}
        \caption{Low-Frequency Set, YZ View}
    \end{subfigure}
    \hfill
    \begin{subfigure}{0.48\textwidth}
        \centering
        \includegraphics[width=\linewidth]{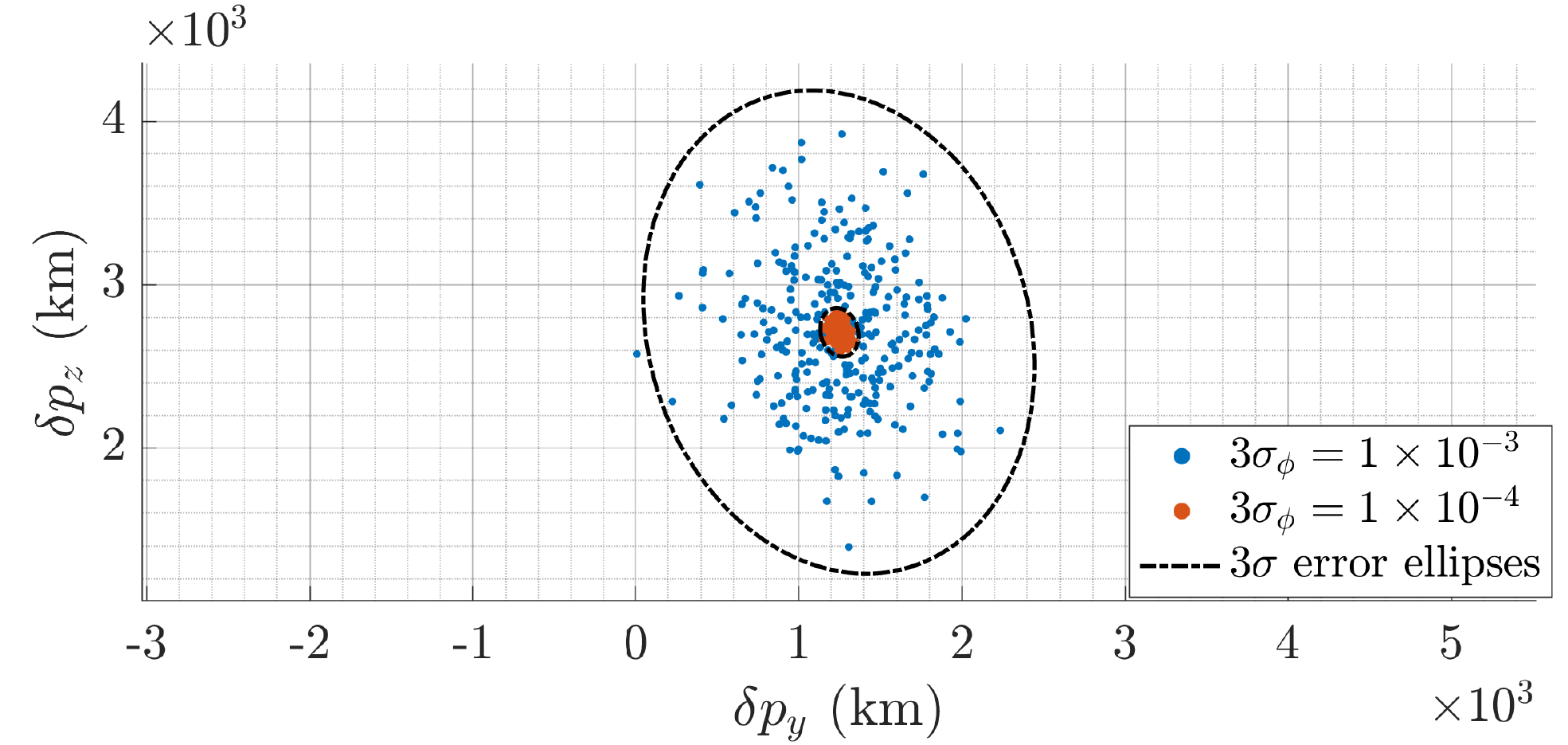}
        \caption{Mixed-Frequency Set, YZ View}
    \end{subfigure}
    \caption{Candidate point error distribution from 300-sample Monte Carlo simulations of the low-frequency pulsar set (left) and the mixed-frequency pulsar set (right) for two phase accuracy values.}
    \label{fig:phase_noise_scatter}
\end{figure}

Higher frequency pulsars or lower phase errors can both reduce the error in the estimated wavefront position. Reducing phase uncertainty by an order of magnitude may require a two-order-of-magnitude increase in measurement time \cite{ray2017characterization}. Depending on mission requirements, it may be beneficial to reduce observation time at the cost of phase accuracy. Figure~\ref{fig:phase_noise_scatter} shows that reducing the phase uncertainty by an order of magnitude leads to nearly an order of magnitude reduction in the dispersion of errors in a 300-sample Monte Carlo simulation. The sample mean is not centered at zero error due to errors in wavefront position caused by time dilation.

Lower frequency pulsars have fewer wavefronts in a given spatial domain, because a greater time of flight $(t_f-t_0)$ is needed to advance the pulsar phase by one integer cycle as shown in Eq.~\eqref{eqn:phase_position}. This means lower frequency pulsars require fewer wavefront intersect computations for the same domain. Choosing pulsars with small angular separation has also been shown to improve computational speed \cite{lohan2022characterization}. In Figure~\ref{fig:phase_noise_scatter}, the mixed-frequency pulsar set provides better accuracy than the low-frequency pulsar set. This is expected, since the mixed-frequency set includes three high frequency SEXTANT pulsars, and higher frequency pulsars have lower wavefront position error for the same phase error, as shown in Eq.~\eqref{eqn:wavefront_error}. When selecting pulsars to observe, the trade-off between computation speed and candidate-point accuracy should be considered.

\subsection{Candidate Point Uniqueness}

\begin{figure}[!h]
    \centering
    \setlength{\belowcaptionskip}{-5pt}
    \includegraphics[width=0.6\linewidth,
                     clip=true,
                     trim={0 1 0 11}]{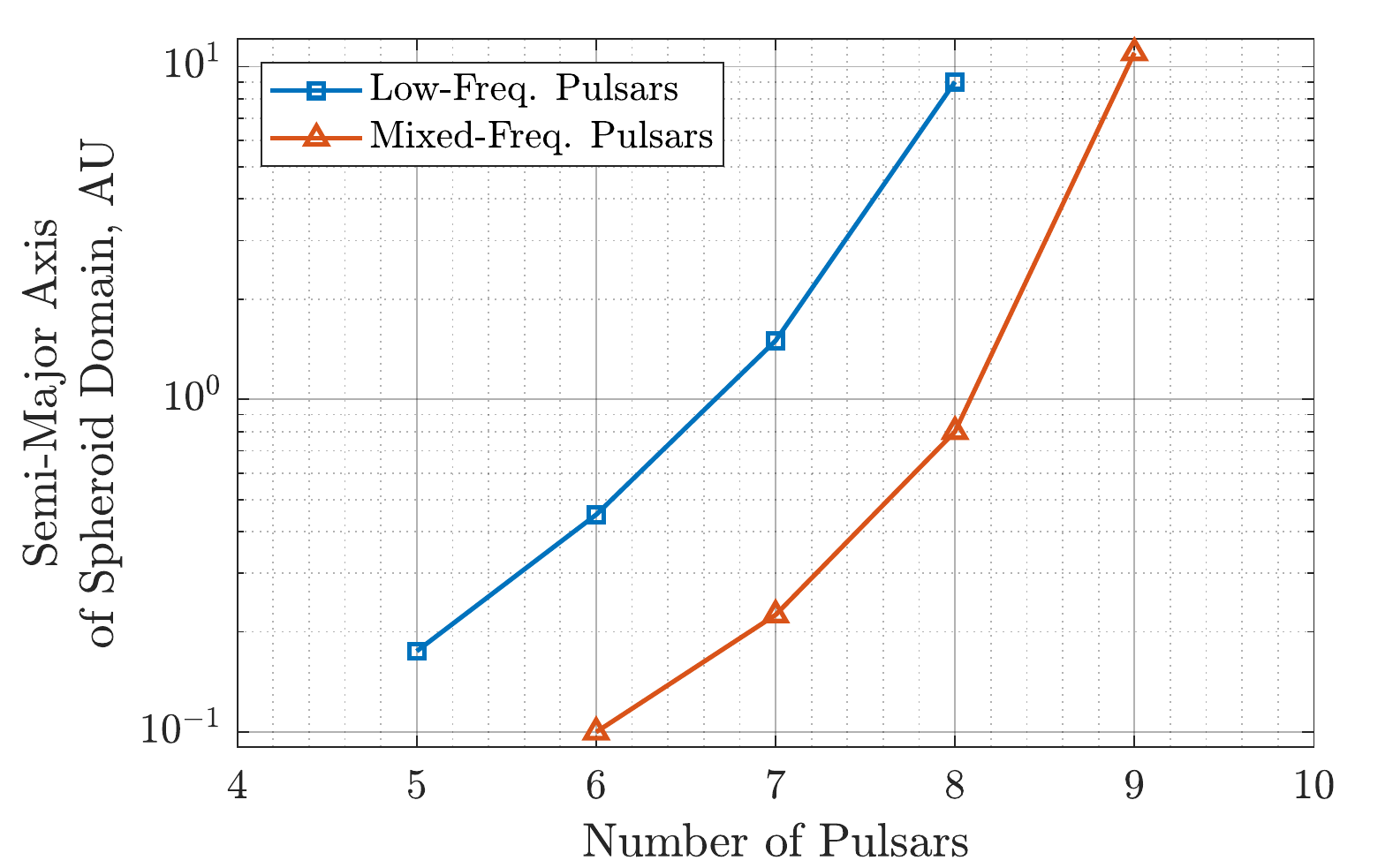}
    \caption{The maximum radius of a spheroid domain in which a unique candidate point can be found, as a function of the number of pulsars observed.}
    \label{fig:numPsr_low_high}
\end{figure}

If an insufficient number of pulsars is observed for a given domain size, the algorithm may identify multiple candidate points, as shown in Figure~\ref{fig:banded_error_model}. In Figure~\ref{fig:numPsr_low_high}, as the number of pulsars used in the algorithm increases, the maximum semi-major axis of the spheroid domain that still yields a unique solution increases. The increase in domain radius is roughly proportional to $\mathcal{O}(\exp{(n^2)})$ with respect to the number of pulsars; additional pulsars dramatically increase the domain size under which a unique solution is found. The two pulsar sets were limited to 9 and 10 pulsars, respectively, due to limitations in computational resources, as their radius limits could be on the order of hundreds of AU if another pulsar is added. Searching such an expansive domain would require days or possibly weeks of computation, because the computation time increases at a rate of $\mathcal{O}(n^3)$ with respect to the semi-major axis of the domain.

\FloatBarrier
\section{Impact of Higher Fidelity Models on Algorithm Performance}
\label{sec:impact-models} 

Solving for candidate positions requires knowledge of the position of pulsar wavefronts, which are estimated through Eq.~\eqref{eqn:get_distance}. Increasing the distance $r$ between the reference point $\mb{b}$ and the true spacecraft position $\mb{p}$ may decrease the accuracy of the estimated delay, which degrades the estimate of wavefront positions. A larger $r$ may lead to reduced accuracy of the candidate points or failure to identify the candidate. 

In practice, this means that some knowledge of the spacecraft true position $\mb{p}$ must be available to be able to place the reference point $\mb{b}$ within some limit. The following analysis determines the upper bound of the reference point distance $r$ by assessing the performance of the algorithm with respect to a range of values of $r$. The impact of higher fidelity models on the upper bound of $r$ is studied by comparing algorithm performance when these models are used in the algorithm.

\FloatBarrier
\subsection{Parallax Effect}

\begin{figure}[!h]
    \centering
    \setlength\belowcaptionskip{-5pt}
    \includegraphics[width=0.6\linewidth]{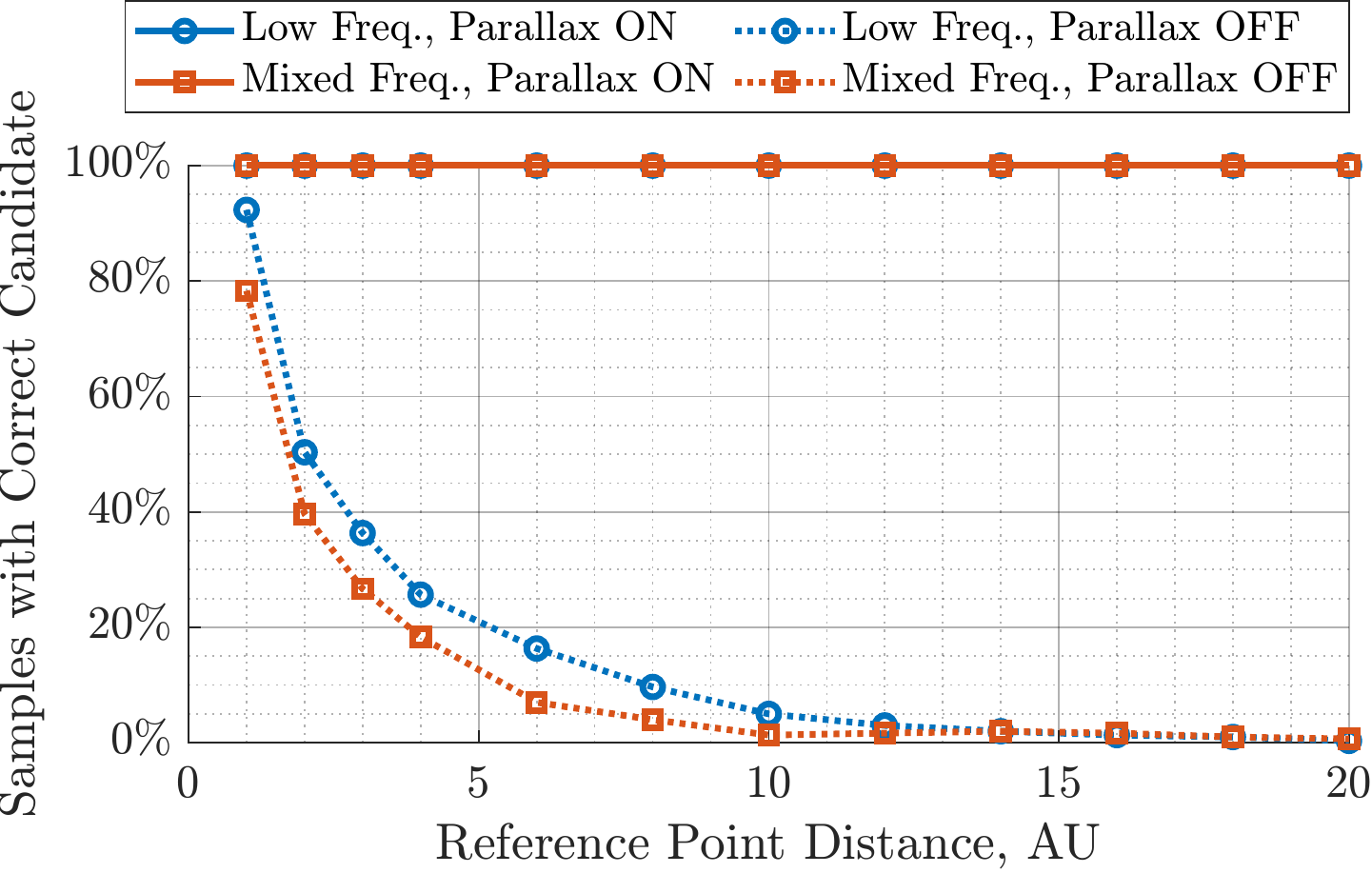}
    \caption{The number of unique, correct candidate point solutions from 300-sample Monte Carlo simulations is shown for two different sets of pulsars with and without parallax effect estimation. Performance degrades significantly for both sets of pulsars when the parallax effect is neglected.}
    \label{fig:parallax_count}
\end{figure}

Figure~\ref{fig:parallax_count} shows the necessity of including the parallax effect when solving for candidate points. When the algorithm is set to neglect the parallax effect, it cannot reliably and correctly identify a candidate point even when the reference point is within 1 AU of the spacecraft true position. When the parallax effect is taken into account within the algorithm, the algorithm correctly identifies the position of the spacecraft in all of the 300 Monte Carlo samples at reference point distances of up to 20 AU. In other words, it is essential to account for the parallax effect when considering XNAV for cold-start applications. Although the low-frequency pulsar set performed better than the mixed-frequency set when the parallax effect was neglected, the two sets had nearly identical performance in the more practical case where the parallax effect is accounted for in the algorithm.

Box-and-whisker plots are used to visualize the distribution of candidate point position error to better capture possibly skewed data. The upper and lower edges of the blue box represent the \nth{25} and \nth{75} percentiles of the sample data; their difference is the interquartile range (IQR). The red line inside the box is the sample median. The black whiskers extend to the farthest values within $1.5\cdot$IQR from the \nth{25} and \nth{75} percentiles. Values beyond the whiskers are considered outliers and are marked by red crosses. The notch captures variability in the sample --- if the notches spanned by two datasets have no overlap on the y-axis, their true median is different at the 5\% significance level. More details can be found in the MATLAB boxplot documentation \cite{MATLABboxplot}.

\begin{figure}[!h]
    \centering
    \begin{subfigure}{0.48\textwidth}
        \centering
        \includegraphics[width=\linewidth]{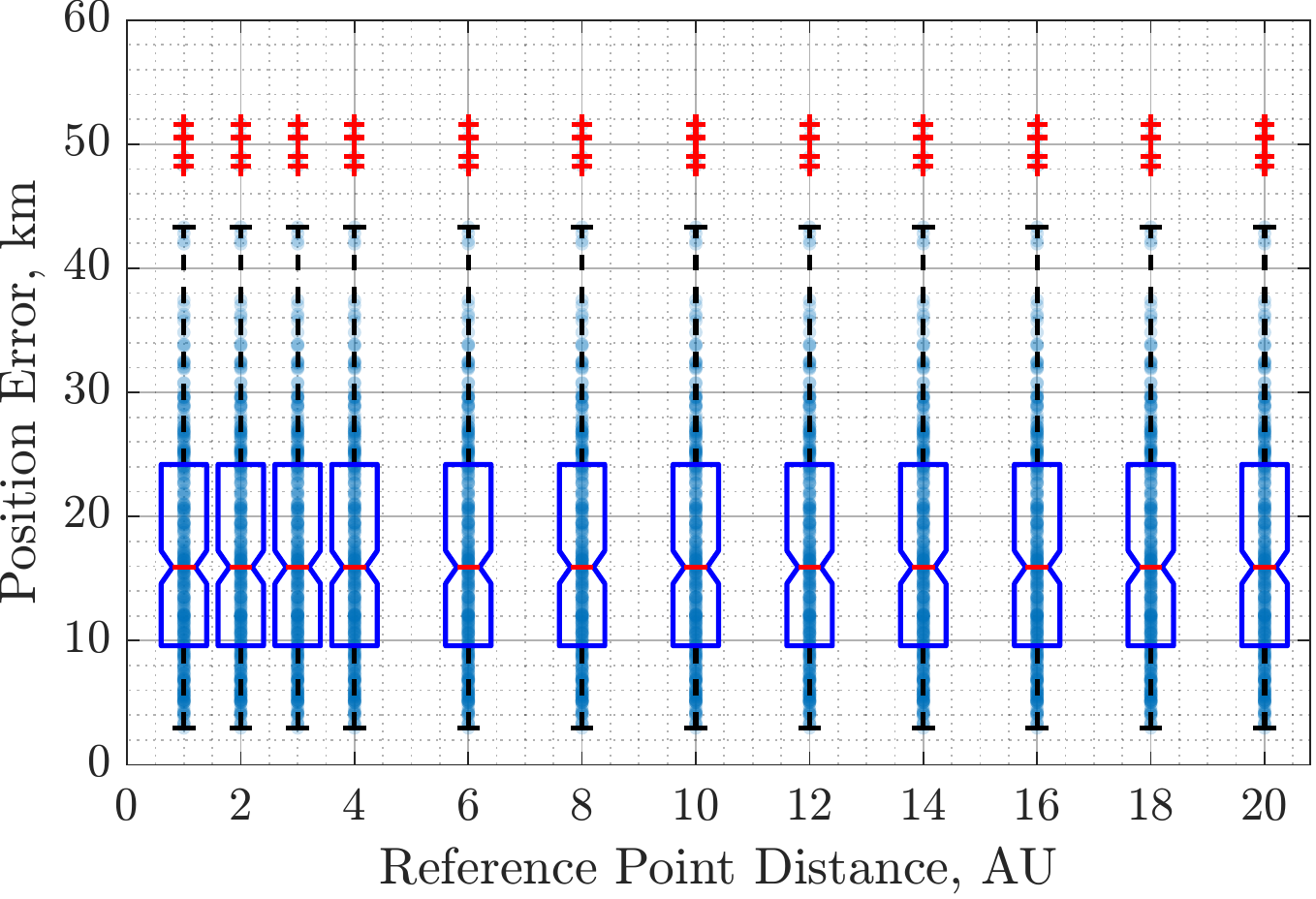}
        \caption{Low-Frequency Pulsars}
    \end{subfigure}
    \hfill
    \begin{subfigure}{0.48\textwidth}
        \centering
        \includegraphics[width=\linewidth]{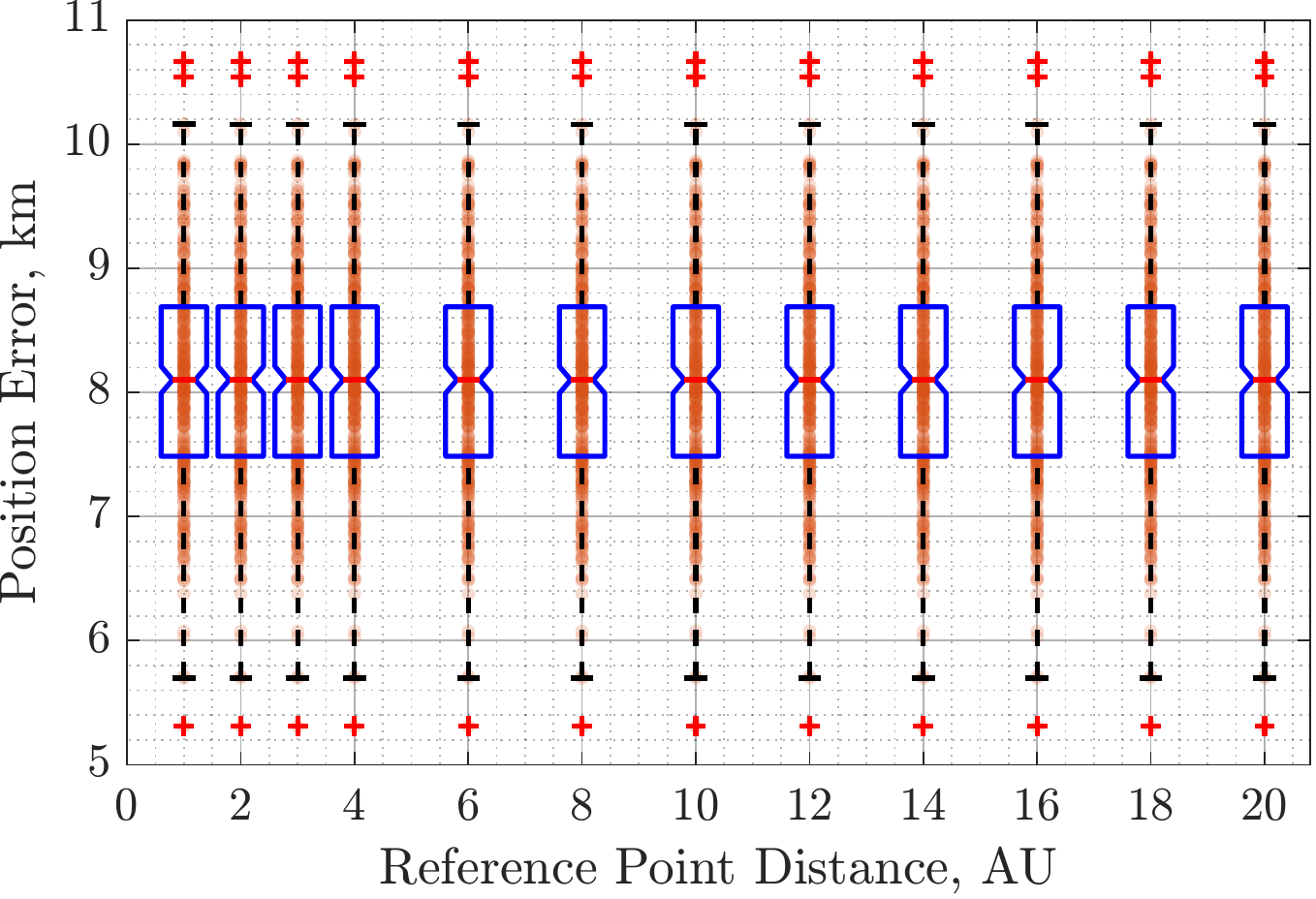}
        \caption{Mixed-Frequency Pulsars}
    \end{subfigure}
\setlength{\belowcaptionskip}{-3pt}
    \caption{The mixed-frequency pulsar set (right) has lower median position error and better precision as compared to the low-frequency pulsar set (left), which is also more skewed.}
    \label{fig:parallax_stats}
\end{figure}

Figure~\ref{fig:parallax_stats} shows that, when the parallax effect is accounted for in the algorithm, the mixed-frequency pulsar set produces better position accuracy compared to the low-frequency set, having a lower median and a smaller IQR. The low-frequency set produced results more prominently skewed toward higher error values. The positioning errors for each pulsar set are nearly identical across different values of reference point distances.

\FloatBarrier
\subsection{Shapiro Delay}

\begin{figure}[!h]
    \centering
    \setlength\belowcaptionskip{-5pt}
    \includegraphics[width=0.6\linewidth]{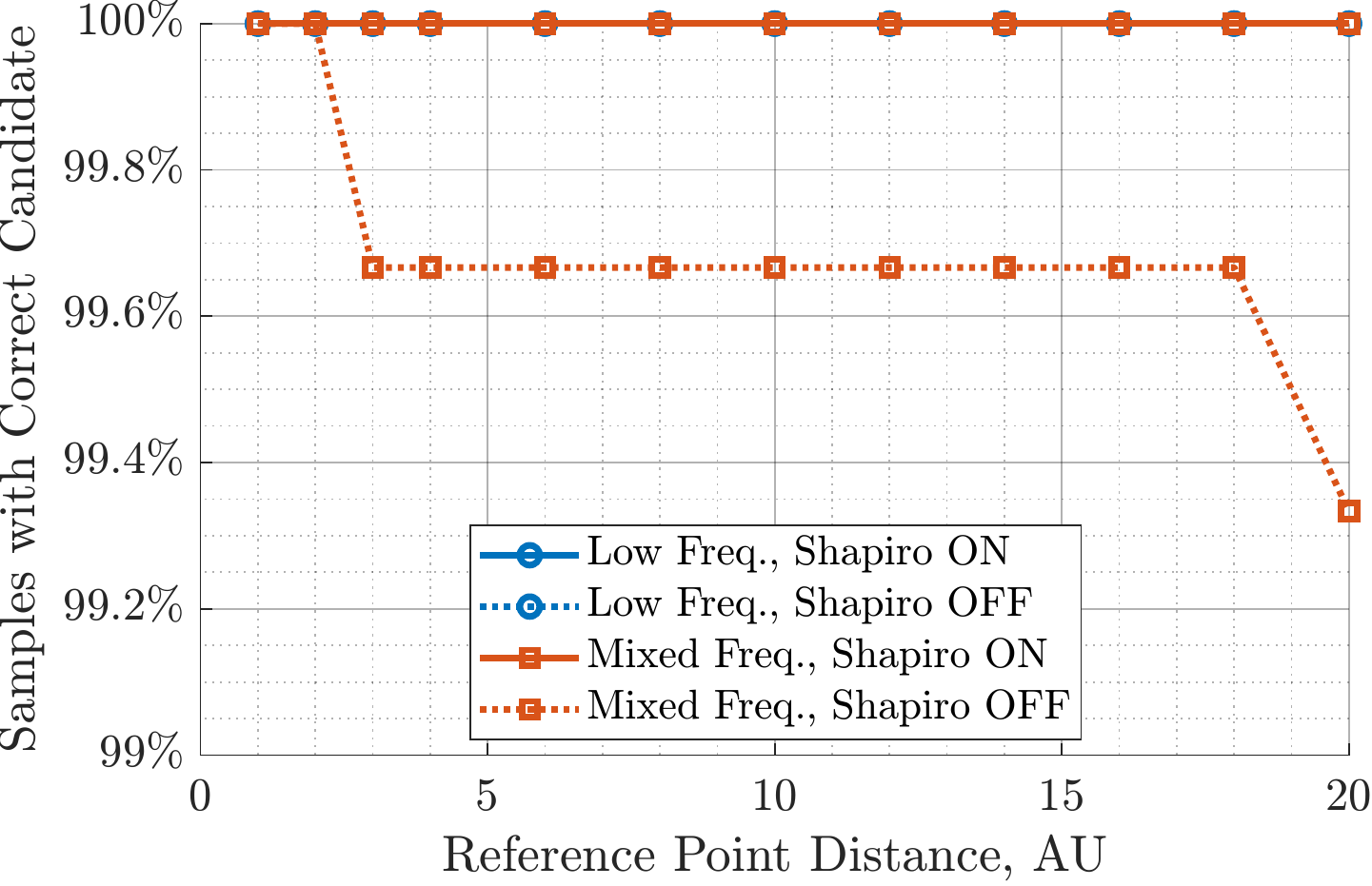}
    \caption{The number of unique, correct candidate point solutions from 300-sample Monte Carlo simulations is performed over a range of reference point distances. The simulations are repeated with Shapiro delay estimation enabled/disabled, and for two different sets of pulsars.}
    \label{fig:shapiro_count}
\end{figure}

Figure~\ref{fig:shapiro_count} shows that the Shapiro delay has little impact on algorithm ability to determine a unique candidate solution. For both sets of pulsars and at all reference point distances, accounting for the Shapiro delay resulted in all Monte Carlo samples having the correct solution. For the mixed-frequency pulsar set, at reference point distances beyond 2 AU, up to two of the 300 samples did not produce a correct solution.

\begin{figure}[!h]
\centering
\begin{subfigure}{0.48\textwidth}
    \centering
    \includegraphics[width=\linewidth]{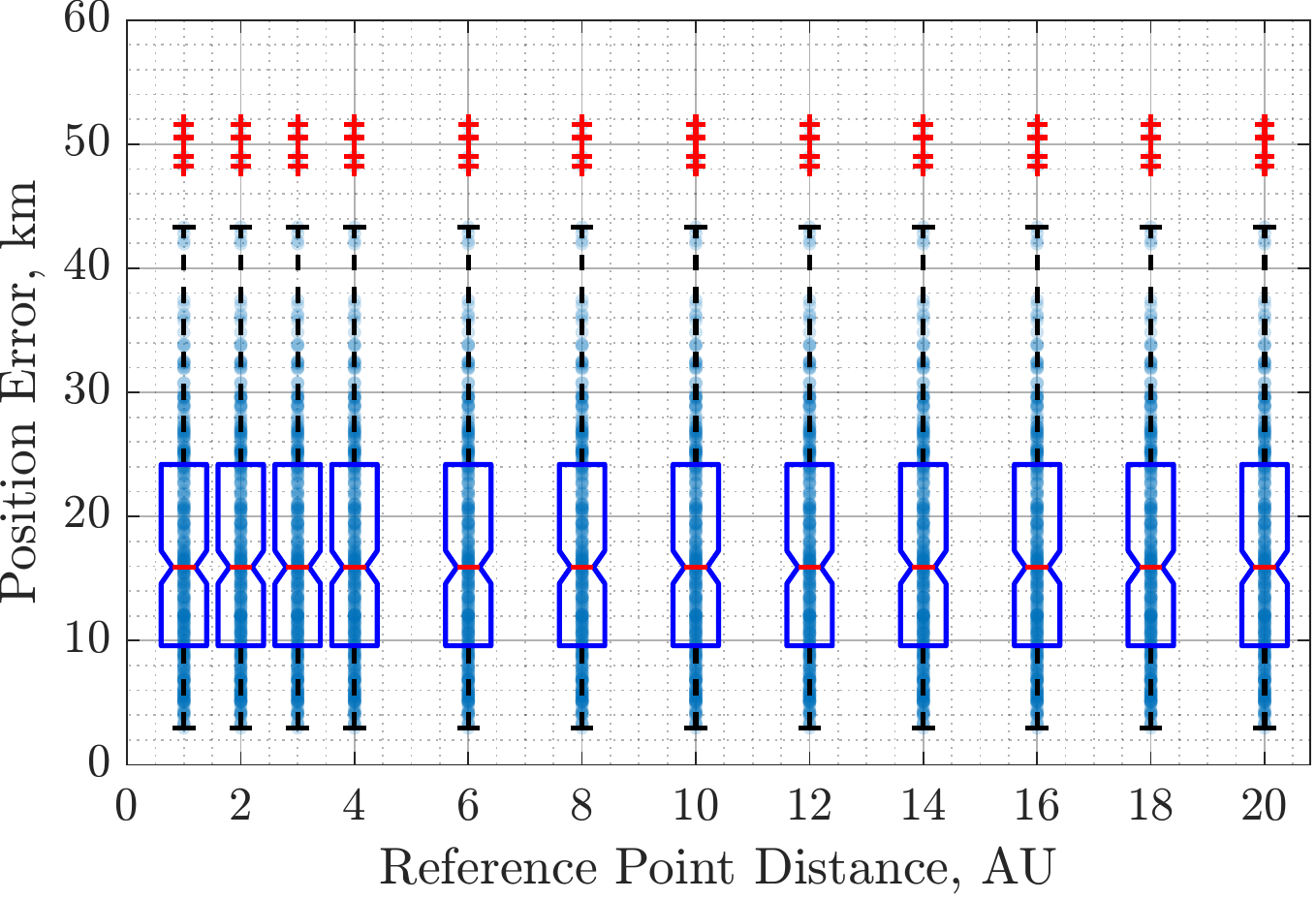}
    \caption{Low Frequency Pulsars, Shapiro Enabled}
\end{subfigure}
\hfill
\begin{subfigure}{0.48\textwidth}
    \centering
    \includegraphics[width=\linewidth]{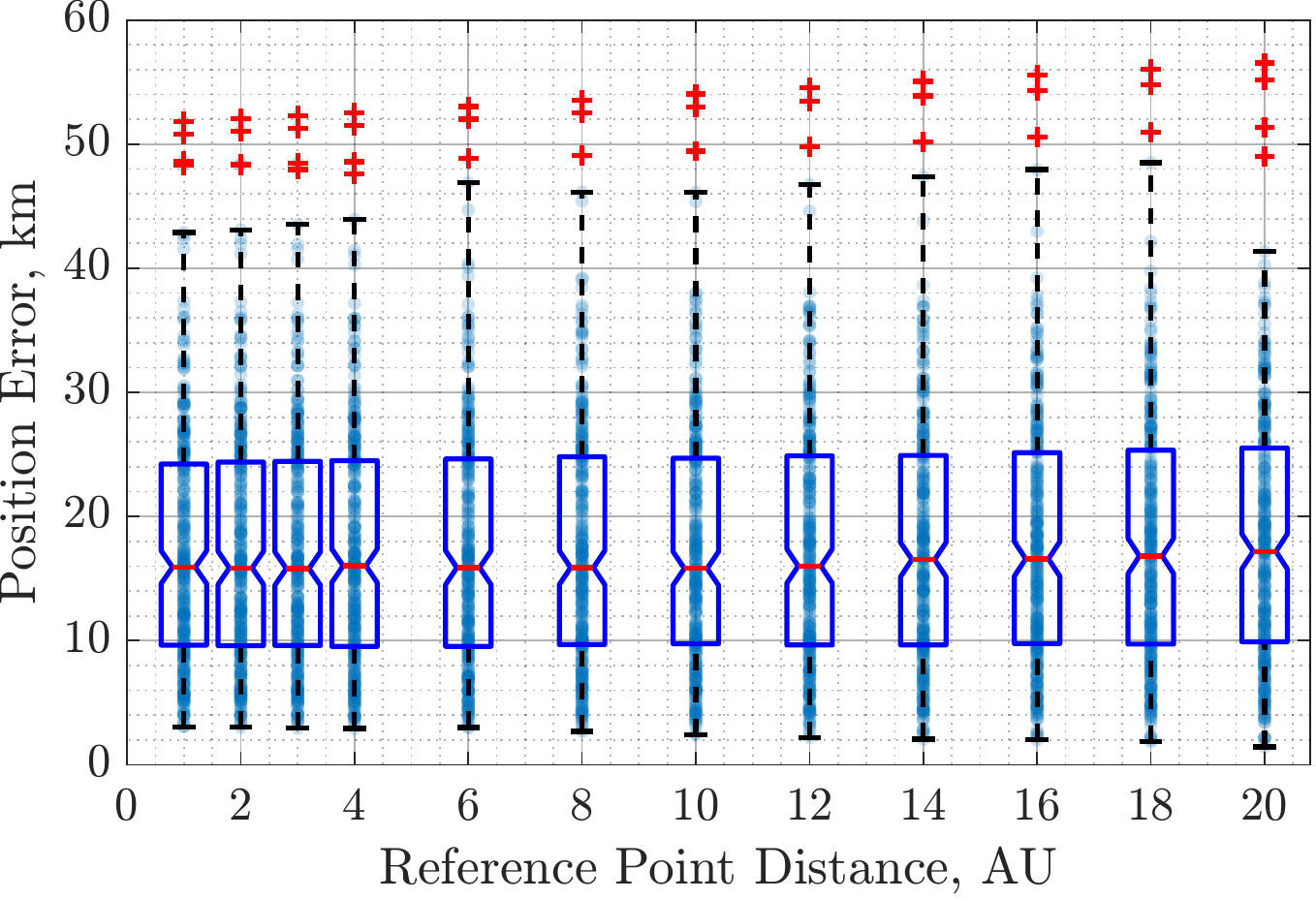}
    \caption{Low Frequency Pulsars, Shapiro Disabled}
\end{subfigure}
\vskip\parskip
\begin{subfigure}{0.48\textwidth}
    \centering
    \includegraphics[width=\linewidth]{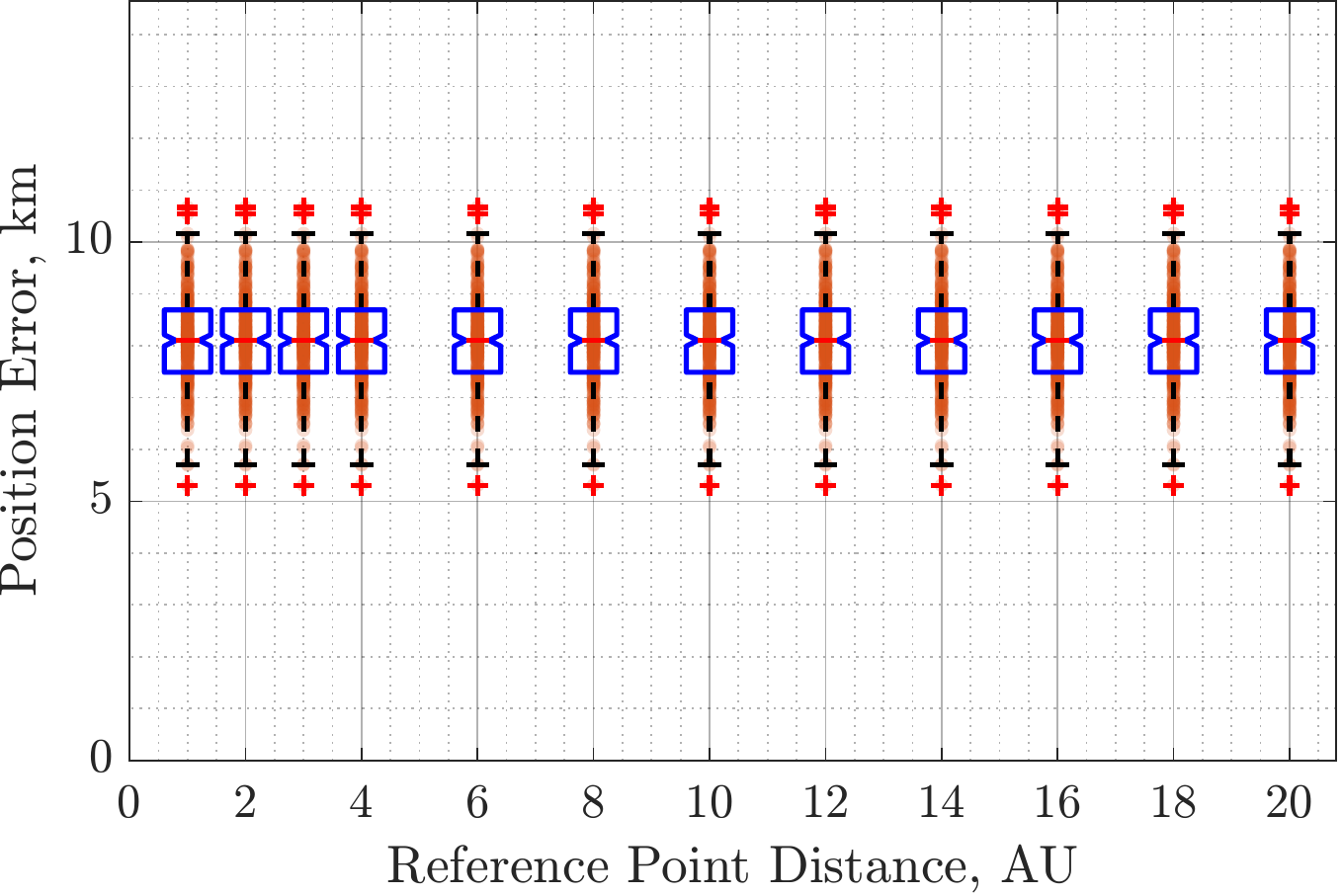}
    \caption{High Frequency Pulsars, Shapiro Enabled}
\end{subfigure}
\hfill
\begin{subfigure}{0.48\textwidth}
    \centering
    \includegraphics[width=\linewidth]{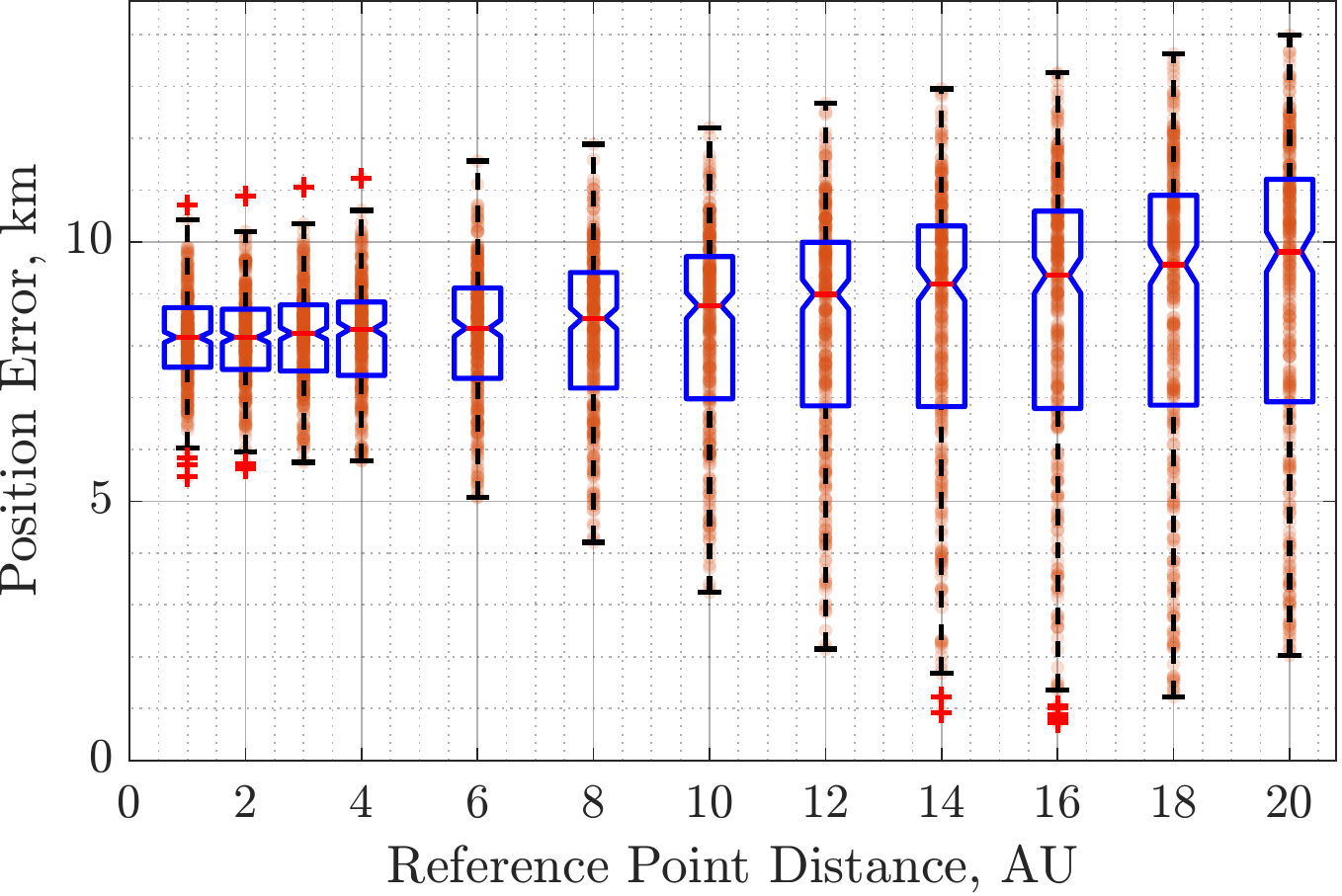}
    \caption{High Frequency Pulsar, Shapiro Disabled}
\end{subfigure}
\caption{Neglecting the Shapiro delay reduced the median position error for both sets of pulsars. Precision was reduced for the mixed-frequency pulsar set as the reference point distance increases when Shapiro delay is neglected.}
\label{fig:shapiro_stats}
\end{figure}

Figure~\ref{fig:shapiro_stats} shows that enabling Shapiro delay estimation produces a more consistent error across all reference point distances. When the algorithm neglects Shapiro delay, both the low- and mixed-frequency pulsar sets had worse median errors that increased with reference point distance, although the degradation of accuracy was much more prominent for the mixed-frequency pulsar set. The mixed-frequency pulsar set also becomes more skewed toward lower error values as the reference point distance increases.

\FloatBarrier
\subsection{Time Dilation}

The spacecraft onboard clock measures proper time, which was found to be approximately 2.5022 \si{\ms} slower than barycentric coordinate time using Eq.~\eqref{eqn:time_dilation_heliocentric} with the chosen mission profile. Since some pulsars used in the mixed-frequency set have frequencies on the order of 100 Hz, their measured phase can be offset by 20\% or more due to time dilation alone. In comparison, studies have previously considered phase accuracy on the order of $10^{-3}$ to $10^{-6}$ \cite{ray2017characterization,sun2018building}. This problem would be further exacerbated if the onboard clock is not synchronized for a longer time period, or if the spacecraft is deeper in a gravitational well. To successfully solve the cold-start problem using XNAV, an estimate of time dilation is necessary to maintain phase measurement accuracy.

One technique for estimating time dilation is propagating the last known spacecraft trajectory using available onboard models to obtain an approximation to the actual time dilation. For example, two-body dynamics may be assumed for a spacecraft in heliocentric orbit. In this case, the relation between proper time $\tau$ and coordinate time $t$ can be directly evaluated using Eq.~\eqref{eqn:time_dilation_heliocentric} \cite{sheikh2006spacecraft}, where $\mu_{\bm{_\odot}}$ is the Sun's gravitational parameter, $E$ is the spacecraft's eccentric anomaly, $c$ is the speed of light in a vacuum, and $a$ is the spacecraft orbit's semi-major axis.
\begin{equation}
    \label{eqn:time_dilation_heliocentric}
    (t-t_0) = (\tau-\tau_0)\Big(1-\dfrac{\mu_{\bm{_\odot}}}{2c^2a}\Big) + \Big(\dfrac{2}{c^2}\Big)\sqrt{a\mu_{\bm{_\odot}}}(E-E_0)
\end{equation}

Using the chosen trajectory, the estimated time dilation assuming two-body dynamics is 2.5020 \si{\ms}, which is only 200 \si{\ns} from the true value. Other techniques for estimating time dilation may also exist, depending on the information available to the spacecraft. However, a detailed discussion of methods to estimate time dilation is beyond the scope of this study. Instead, the performance of the algorithm is evaluated with respect to a range of time dilation estimate errors to assess the estimation accuracy required to solve the cold-start problem.

\begin{figure}[ht!]
    \centering
    \includegraphics[width=0.6\linewidth]{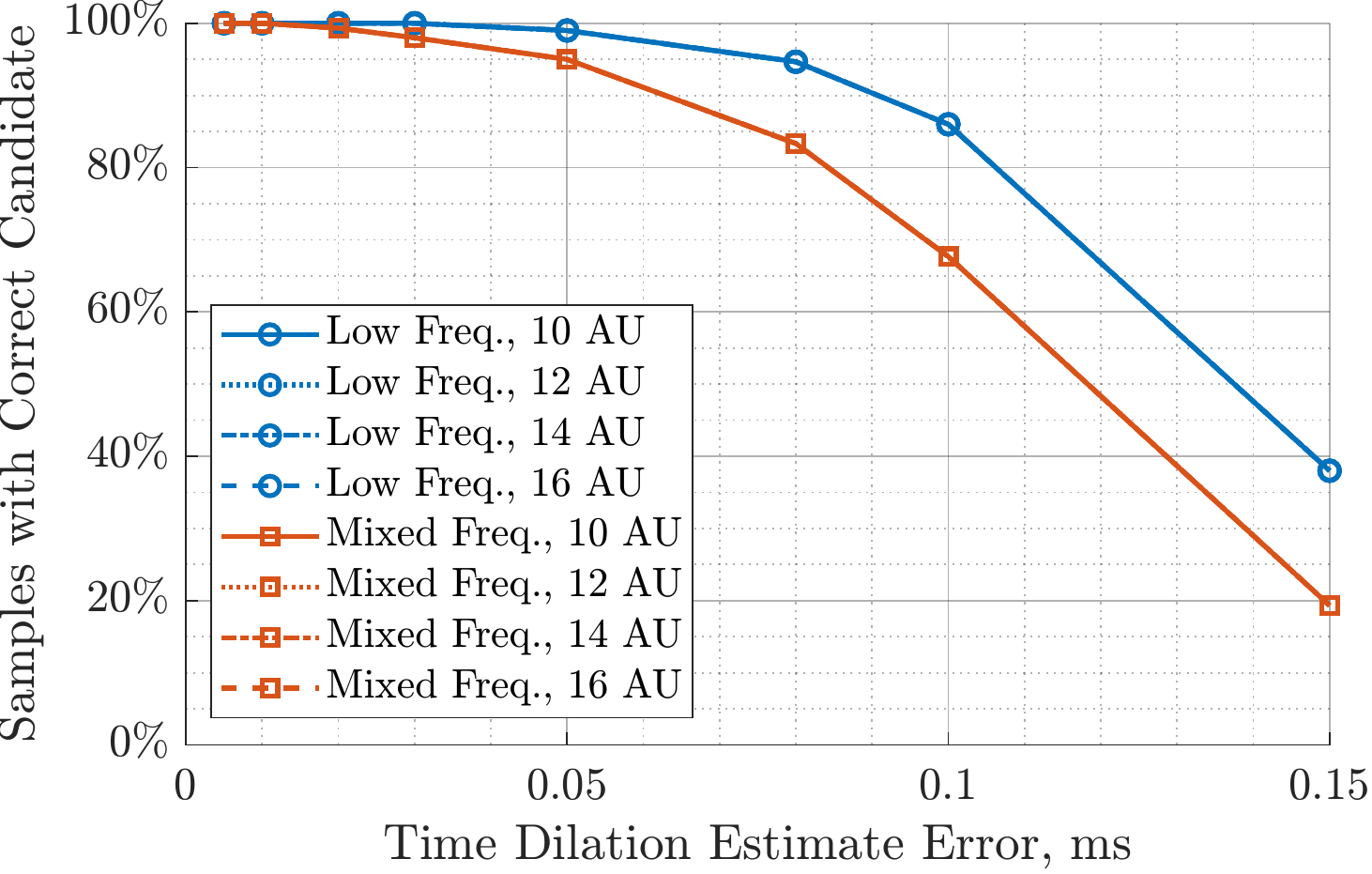}
    \caption{The number of unique, correct candidate point solutions decreases as the time dilation estimate error increases. The reference point distance does not affect the number of correct solutions; the low-frequency pulsar set has more correct solutions than the mixed-frequency set for the same time dilation estimate error.}
    \label{fig:einstein_count}
\end{figure}

Figure~\ref{fig:einstein_count} shows that the algorithm is less likely to identify a unique and correct solution as the error in the time dilation estimate increases. Furthermore, this degradation appears to be independent of the reference point distance for both pulsar sets. 

\begin{figure}[ht!]
    \centering
    \begin{subfigure}{0.48\textwidth}
        \centering
        \includegraphics[width=\linewidth]{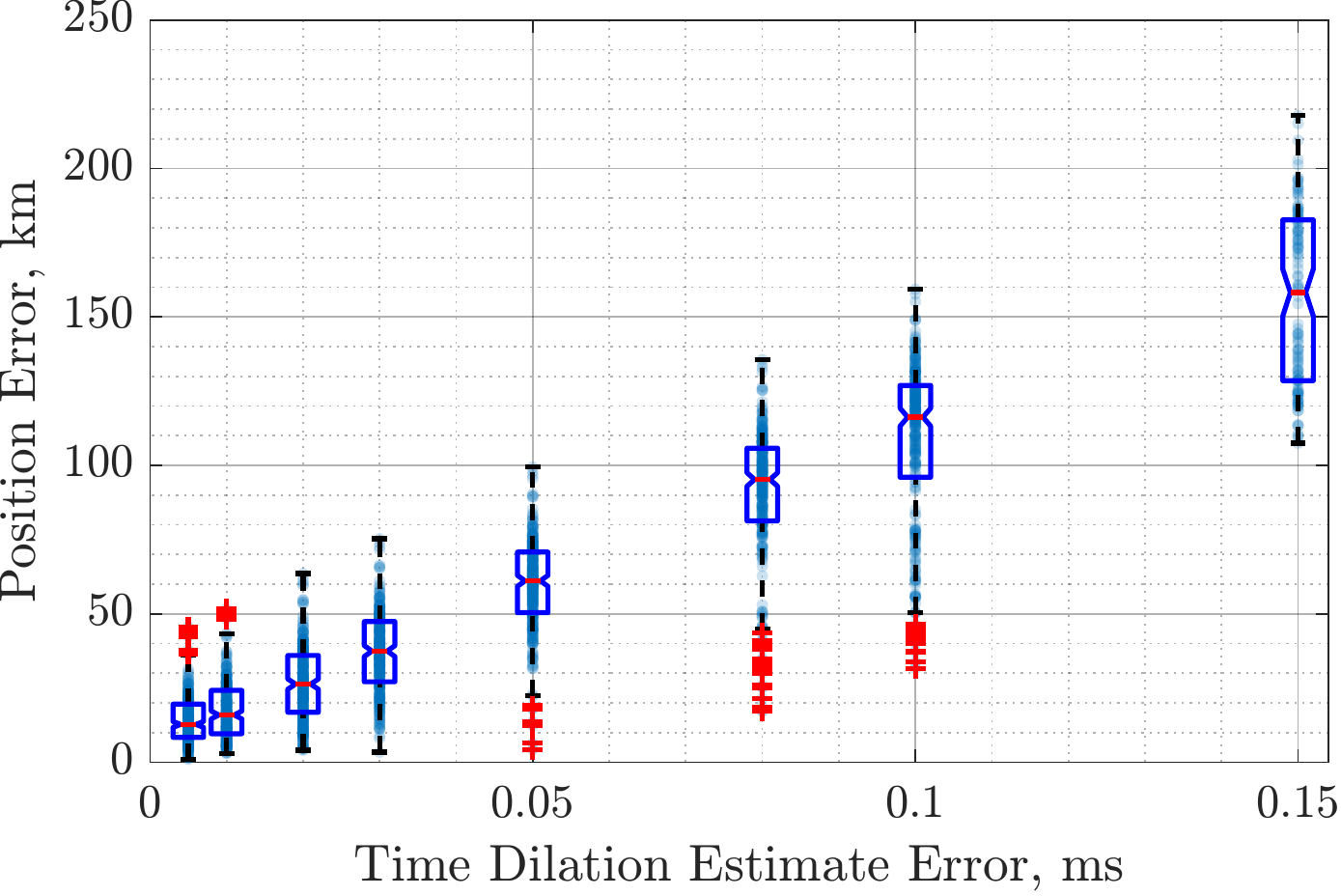}
        \caption{Low-Frequency Pulsar Set}
    \end{subfigure}
    \hfill
    \begin{subfigure}{0.48\textwidth}
        \centering
        \includegraphics[width=\linewidth]{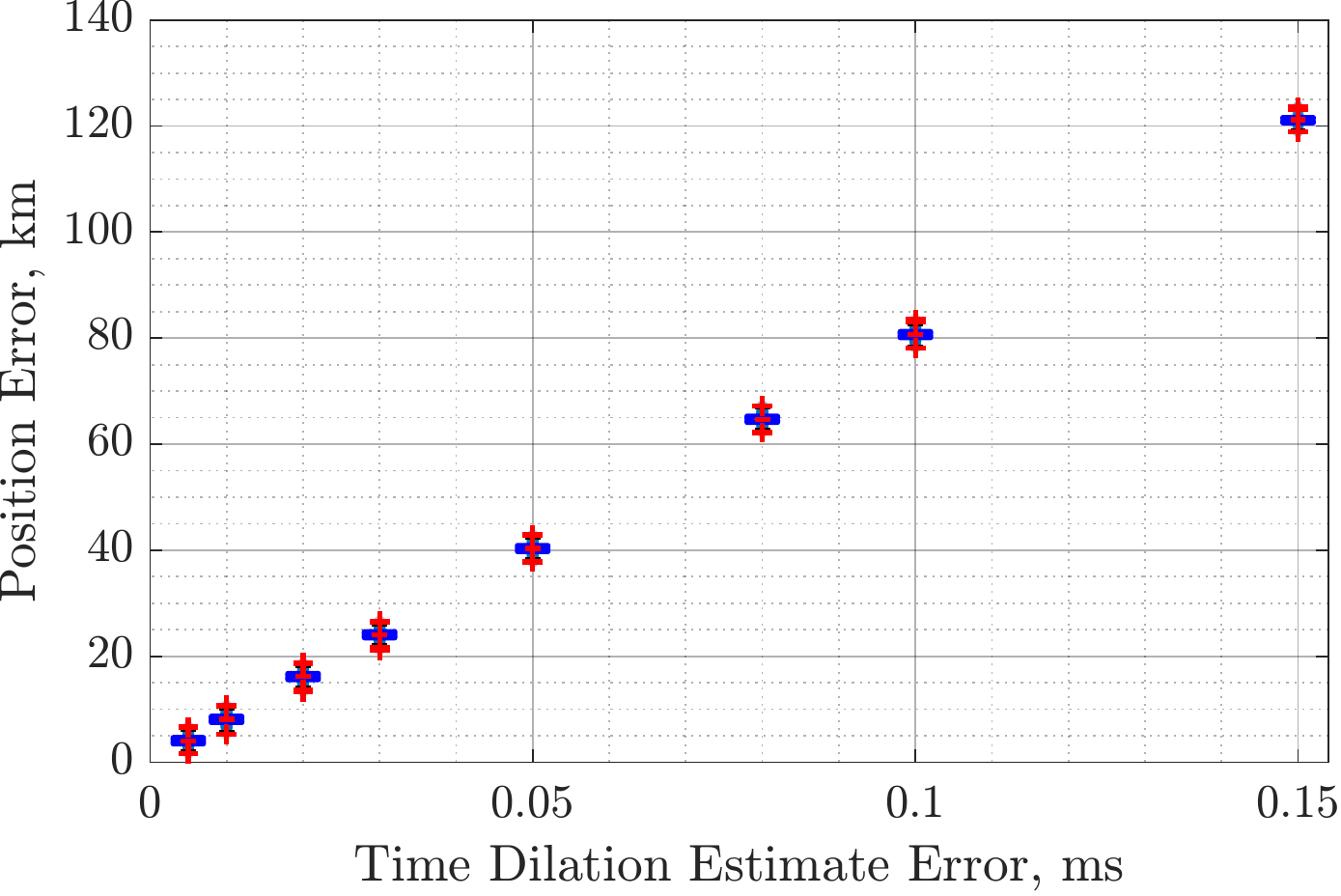}
        \caption{Mixed-Frequency Pulsar Set}
    \end{subfigure}
    \setlength{\belowcaptionskip}{-10pt}
    \caption{The candidate point position error of correct solutions for both the low-frequency pulsar set (left) and the mixed-frequency pulsar set (right) increase linearly with respect to time dilation estimate error. The IQR increases with greater time dilation estimate error.}
    \label{fig:einstein_stats}
\end{figure}

Figure~\ref{fig:einstein_stats} shows the position error of correct candidate solutions at 10 AU. Results for other reference point distance values are not shown because they are nearly identical, since the reference point distance has no significant impact on the effects of the time dilation estimate. There is a positive correlation between the time dilation estimate error and the candidate point position error. In particular, for the mixed-frequency pulsar set, the trend appears to be linear. This is expected behavior --- since the time dilation estimate error is much less than one second, changes in pulsar frequency are negligible, making phase $\phi$ linear with respect to time $t$ as in Eq.~\eqref{eqn:phase_both}. Therefore, the errors in $t$ are linearly related to the errors in $\phi$, which in turn influences candidate position error.

Comparing Figure~\ref{fig:einstein_count} and Figure~\ref{fig:einstein_stats} reveals that the success rate of the algorithm is the limiting factor in the accuracy of the required time dilation estimate. Although the candidate point position error increases roughly linearly even with a time dilation estimate error of 0.15 \si{\ms}, the time dilation estimate error should be kept under 70 \si{\us} and 50 \si{\us} for the low- and mixed-frequency pulsar sets to correctly identify candidate solutions in at least 95\% of cases.

\section{Conclusion}
\label{sec:conclusion}

A norm-minimization-based algorithm is proposed to solve the cold-start problem using XNAV. The Shapiro delay, parallax effect, time dilation, and pulsar frequency variations are incorporated into the algorithm in addition to the first order Doppler delay considered by existing studies. Results indicate that the parallax effect must be taken into account when using the algorithm to solve the cold-start problem and that time dilation must be estimated to within 50-70 \si{\us} for reliable convergence, depending on the pulsars observed. However, accounting for the Shapiro delay appears to have a negligible impact on algorithm performance. 

The proposed norm-minimization algorithm is a new contribution toward solutions to the cold-start problem. The algorithm can uniquely identifying a point in space within 5-100 km of the true spacecraft position given simultaneous pulsar phase observations. Varying levels of accuracy and specificity can be achieved depending on the quality of observations and the pulsars observed.

Additional research into methods for processing sequential pulsar observations may improve the applicability of the algorithm. Further research into pulsar candidate selection may enable more intelligent pulsar selection based on the desired metric, such as minimizing error, reducing the number of candidate points, or reducing computation time.

\FloatBarrier


\bibliographystyle{AAS_publication}   
\bibliography{references}   

\begin{thebibliography}{10}

\bibitem{allan1987millisecond}
D.~Allan, ``Millisecond Pulsar Rivals Best Atomic Clock Stability,''  {\em 41st
  Annual Symposium on Frequency Control}, IEEE, 1987, pp.~2--11,
  \doi{10.1109/freq.1987.200994}.

\bibitem{graven2008xnav}
P.~Graven, J.~Collins, S.~Sheikh, J.~Hanson, P.~Ray, and K.~Wood, ``XNAV for
  Deep Space Navigation,''  {\em 31st Annual AAS Guidance and Control
  Conference}, AAS San Diego, 2008.
\newblock AAS 08-054.

\bibitem{winternitz2016sextant}
L.~M. Winternitz, J.~W. Mitchell, M.~A. Hassouneh, J.~E. Valdez, S.~R. Price,
  S.~R. Semper, H.~Y. Wayne, P.~S. Ray, K.~S. Wood, Z.~Arzoumanian, {\em
  et~al.}, ``SEXTANT X-ray Pulsar Navigation Demonstration: Flight System and
  Test Results,''  {\em 2016 IEEE Aerospace Conference}, IEEE, 2016,
  \doi{10.1109/AERO.2016.7500838}.

\bibitem{tanygin2014closed}
S.~Tanygin, ``Closed-form Solution for Lost-in-Space Visual Navigation
  Problem,''  {\em Journal of Guidance, Control, and Dynamics}, Vol.~37, No.~6,
  2014, pp.~1754--1766, \doi{10.2514/1.G000529}.

\bibitem{adams2017lost}
V.~H. Adams and M.~A. Peck, ``Lost in Space and Time,''  {\em AIAA Guidance,
  Navigation, and Control Conference}, 2017, p.~1030,
  \doi{10.2514/6.2017-1030}.

\bibitem{dahir2020lost}
A.~R. Dahir, {\em Lost in Space: Autonomous Deep Space Navigation}.
\newblock PhD thesis, University of Colorado at Boulder, 2020.

\bibitem{hollenberg2020geometric}
C.~L. Hollenberg and J.~A. Christian, ``Geometric Solutions for Problems in
  Velocity-Based Orbit Determination,''  {\em The Journal of the Astronautical
  Sciences}, Vol.~67, No.~1, 2020, pp.~188--224,
  \doi{10.1007/s40295-019-00170-7}.

\bibitem{hou2021compare}
L.~Hou, K.~Lohan, and Z.~R. Putnam, ``Comparison and Error Modeling of
  Velocity-Based Initial Orbit Determination Algorithms,''  {\em AAS/AIAA Space
  Flight Mechanics Meeting}, Virtual, February 2021.
\newblock AAS 21-280.

\bibitem{sheikhUseVariableCelestial2005}
S.~I. Sheikh, {\em The Use Of Variable Celestial X-ray Sources For Spacecraft
  Navigation}.
\newblock PhD thesis, University of Maryland, Sept. 2005.
\newblock uri: \url{http://hdl.handle.net/1903/2856}.

\bibitem{lohan2021methodology}
K.~Lohan, {\em Methodology for State Determination without Prior Information
  using X-ray Pulsar Navigation Systems}.
\newblock PhD thesis, University of Illinois at Urbana-Champaign, 2021.

\bibitem{sheikh2007high}
S.~I. Sheikh, R.~W. Hellings, and R.~A. Matzner, ``High-Order Pulsar Timing for
  Navigation,''  {\em Proceedings of the 63rd Annual Meeting of The Institute
  of Navigation}, 2007, pp.~432--443.

\bibitem{ATNFPulsarCatalogue}
G.~Hobbs, R.~Manchester, A.~Teoh, and M.~Hobbs, ``The ATNF Pulsar Catalog,''
  {\em Symposium-International Astronomical Union}, Vol.~218, Cambridge
  University Press, 2004, pp.~139--140, \doi{10.1017/s0074180900180829}.

\bibitem{huygens1899oeuvres}
C.~Huygens, {\em Oeuvres Compl{\`e}tes de Christiaan Huygens}, Vol.~8.
\newblock M. Nijhoff, 1899, \doi{10.5962/bhl.title.21031}.

\bibitem{edwards2006tempo2}
R.~T. Edwards, G.~Hobbs, and R.~Manchester, ``TEMPO2, a new pulsar timing
  package -- II. The timing model and precision estimates,''  {\em Monthly
  Notices of the Royal Astronomical Society}, Vol.~372, No.~4, 2006,
  pp.~1549--1574, \doi{10.1111/j.1365-2966.2006.10870.x}.

\bibitem{backer1986pulsar}
D.~Backer and R.~Hellings, ``Pulsar Timing and General Relativity,''  {\em
  Annual Review of Astronomy and Astrophysics}, Vol.~24, No.~1, 1986,
  pp.~537--575, \doi{10.1146/annurev.aa.24.090186.002541}.

\bibitem{desvignes2016high}
G.~Desvignes, R.~Caballero, L.~Lentati, J.~Verbiest, D.~Champion, B.~Stappers,
  G.~Janssen, P.~Lazarus, S.~Os{\l}owski, S.~Babak, {\em et~al.},
  ``High-precision timing of 42 millisecond pulsars with the European Pulsar
  Timing Array,''  {\em Monthly Notices of the Royal Astronomical Society},
  Vol.~458, No.~3, 2016, pp.~3341--3380, \doi{10.1093/mnras/stw483}.

\bibitem{freire2012relativistic}
P.~C. Freire, N.~Wex, G.~Esposito-Far{\`e}se, J.~P. Verbiest, M.~Bailes, B.~A.
  Jacoby, M.~Kramer, I.~H. Stairs, J.~Antoniadis, and G.~H. Janssen, ``The
  relativistic pulsar--white dwarf binary PSR J1738+ 0333--II. The most
  stringent test of scalar--tensor gravity,''  {\em Monthly Notices of the
  Royal Astronomical Society}, Vol.~423, No.~4, 2012, pp.~3328--3343,
  \doi{10.1111/j.1365-2966.2012.21253.x}.

\bibitem{thomas1975reformulation}
J.~Thomas, ``Reformulation of the relativistic conversion between coordinate
  time and atomic time,''  {\em The Astronomical Journal}, Vol.~80, 1975,
  pp.~405--411, \doi{10.1086/111756}.

\bibitem{moyer1981transformation}
T.~D. Moyer, ``Transformation from proper time on Earth to coordinate time in
  solar system barycentric space-time frame of reference,''  {\em Celestial
  Mechanics}, Vol.~23, No.~1, 1981, pp.~33--56, \doi{10.1007/BF01228543}.

\bibitem{emadzadeh2011x}
A.~A. Emadzadeh and J.~L. Speyer, ``X-ray pulsar-based relative navigation
  using epoch folding,''  {\em IEEE transactions on Aerospace and Electronic
  Systems}, Vol.~47, No.~4, 2011, pp.~2317--2328,
  \doi{0.1109/TAES.2011.6034635}.

\bibitem{zhang2014new}
H.~Zhang, L.-p. Xu, Y.-h. Shen, R.~Jiao, and J.-r. Sun, ``A new
  maximum-likelihood phase estimation method for X-ray pulsar signals,''  {\em
  Journal of Zhejiang University SCIENCE C}, Vol.~15, No.~6, 2014,
  pp.~458--469, \doi{10.1631/jzus.C1300347}.

\bibitem{SPICECoordOverview}
``An Overview of Reference Frame and Coordinate Systems in the SPICE Context,''
   \url{https://naif.jpl.nasa.gov/pub/naif/toolkit_docs/Tutorials/pdf/individual_docs/17_frames_and_coordinate_systems.pdf}.
\newblock Accessed: 2022-03-30.

\bibitem{GSFCTimingSystems}
``The ABC of XTE {\textbar} A Time Tutorial,''
  \url{https://heasarc.gsfc.nasa.gov/docs/xte/abc/time_tutorial.html}.
\newblock Accessed: 2022-03-30.

\bibitem{gmat}
NASA, ``General Mission Analysis Tool (GMAT),''
\newblock \url{https://sourceforge.net/projects/gmat/}.

\bibitem{shemar2016towards}
S.~Shemar, G.~Fraser, L.~Heil, D.~Hindley, A.~Martindale, P.~Molyneux, J.~Pye,
  R.~Warwick, and A.~Lamb, ``Towards practical autonomous deep-space navigation
  using X-Ray pulsar timing,''  {\em Experimental Astronomy}, Vol.~42, No.~2,
  2016, pp.~101--138, \doi{doi.org/10.1007/s10686-016-9496-z}.

\bibitem{ray2017characterization}
P.~S. Ray, K.~S. Wood, and M.~T. Wolff, ``Characterization of pulsar sources
  for x-ray navigation,''  {\em arXiv preprint arXiv:1711.08507}, 2017,
  \doi10.48550/arXiv.1711.08507.

\bibitem{lohan2022characterization}
K.~Lohan and Z.~Putnam, ``Characterization of Candidate Solutions for X-Ray
  Pulsar Navigation,''  {\em IEEE Transactions on Aerospace and Electronic
  Systems}, 2022, \doi{10.1109/TAES.2022.3152684}.

\bibitem{MATLABboxplot}
``Visualize summary statistics with box plot - MATLAB boxplot,''
  \url{https://www.mathworks.com/help/stats/boxplot.html}.
\newblock Accessed: 2022-06-12.

\bibitem{sun2018building}
H.~Sun, X.~Sun, H.~Fang, L.~Shen, S.~Cong, Y.~Liu, X.~Li, and W.~Bao,
  ``Building X-ray pulsar timing model without the use of radio parameters,''
  {\em Acta Astronautica}, Vol.~143, 2018, pp.~155--162,
  \doi{10.1016/j.actaastro.2017.11.014}.

\bibitem{sheikh2006spacecraft}
S.~I. Sheikh, D.~J. Pines, P.~S. Ray, K.~S. Wood, M.~N. Lovellette, and M.~T.
  Wolff, ``Spacecraft Navigation Using X-Ray Pulsars,''  {\em Journal of
  Guidance, Control, and Dynamics}, Vol.~29, No.~1, 2006, pp.~49--63,
  \doi{10.2514/1.13331}.

\end{thebibliography}

\end{document}